\documentclass[
 reprint,
superscriptaddress,
 amsmath,
 amssymb,
 aps,
longbibliography
]{revtex4-2}

\usepackage[pdftex,colorlinks=true,linkcolor=blue,bookmarks=false,citecolor=blue,urlcolor=blue,hypertexnames=false]{hyperref} 
\usepackage{graphicx}
\usepackage{dcolumn}
\usepackage{bm}
\usepackage{siunitx} 
\usepackage{hyperref}
\usepackage{float}
\usepackage{xcolor}
\usepackage{comment}
\usepackage{upgreek}
\usepackage{setspace}

\definecolor{crimson}{rgb}{0.7, 0.1, 0.1}

\DeclareMathOperator{\tr}{Tr}

\begin{document}

\title{Anomalous localization and multifractality in a kicked quasicrystal}

\author{Toshihiko Shimasaki}
\author{Max Prichard}
\thanks{Equal contribution.}
\author{H. Esat Kondakci}
\thanks{Equal contribution.}
\author{Jared E. Pagett}
\thanks{Equal contribution.}
\author{Yifei Bai}
\author{Peter Dotti}
\author{Alec Cao}
\affiliation{Department of Physics, University of California, Santa Barbara, California 93106, USA}
\author{Tsung-Cheng Lu}
\affiliation{Perimeter Institute for Theoretical Physics, Waterloo, Ontario N2L 2Y5, Canada}
\author{Tarun Grover}
\affiliation{Department of Physics, University of California at San Diego, La Jolla, California 92093, USA}
\author{David M. Weld}
\email{Email: weld@ucsb.edu.}
\affiliation{Department of Physics, University of California, Santa Barbara, California 93106, USA}
\maketitle

\textbf{Multifractal states offer a ``third way'' for quantum matter, neither fully localized nor ergodic, exhibiting singular continuous spectra, self-similar wavefunctions, and transport and entanglement scaling exponents intermediate between extended and localized states~\cite{d2016quantum, abanin2019colloquium,gogolin2016equilibration, mori2018thermalization}. While multifractality in equilibrium systems generally requires fine-tuning to a critical point~\cite{evers2008anderson,richardella2010visualizing,chabe2008experimental}, externally driven quantum matter~\cite{Weitenberg:2021wb} can exhibit multifractal states with no equilibrium counterpart. We report the experimental observation of multifractal matter and anomalous localization in a kicked Aubry-Andr\'e-Harper quasicrystal~\cite{aubry1980analyticity,harpermodel}. Our cold-atom realization of this previously-unexplored model is enabled by apodized Floquet engineering techniques which expand the accessible phase diagram by five orders of magnitude. This kicked quantum quasicrystal exhibits a rich phase diagram~\cite{artuso1992phase, artuso1992fractal, leboeuf_phase-space_1990,Qin_Yin_Chen_2014,cadez_Mondaini_Sacramento_2017} including not only fully localized and fully delocalized phases but also an extended region comprising an intricate nested pattern of localized, delocalized, and multifractal states~\cite{prosen2001dimer}. Mapping transport properties throughout the phase diagram, we observe disorder-driven re-entrant delocalization and sub-ballistic transport, and present a theoretical explanation of these phenomena based on eigenstate multifractality.  These results open up the exploration of new states of matter characterized by an intricate interplay of fractal structure and quantum dynamics.} 

\begin{figure*}[bht!]
    \includegraphics[scale=1]{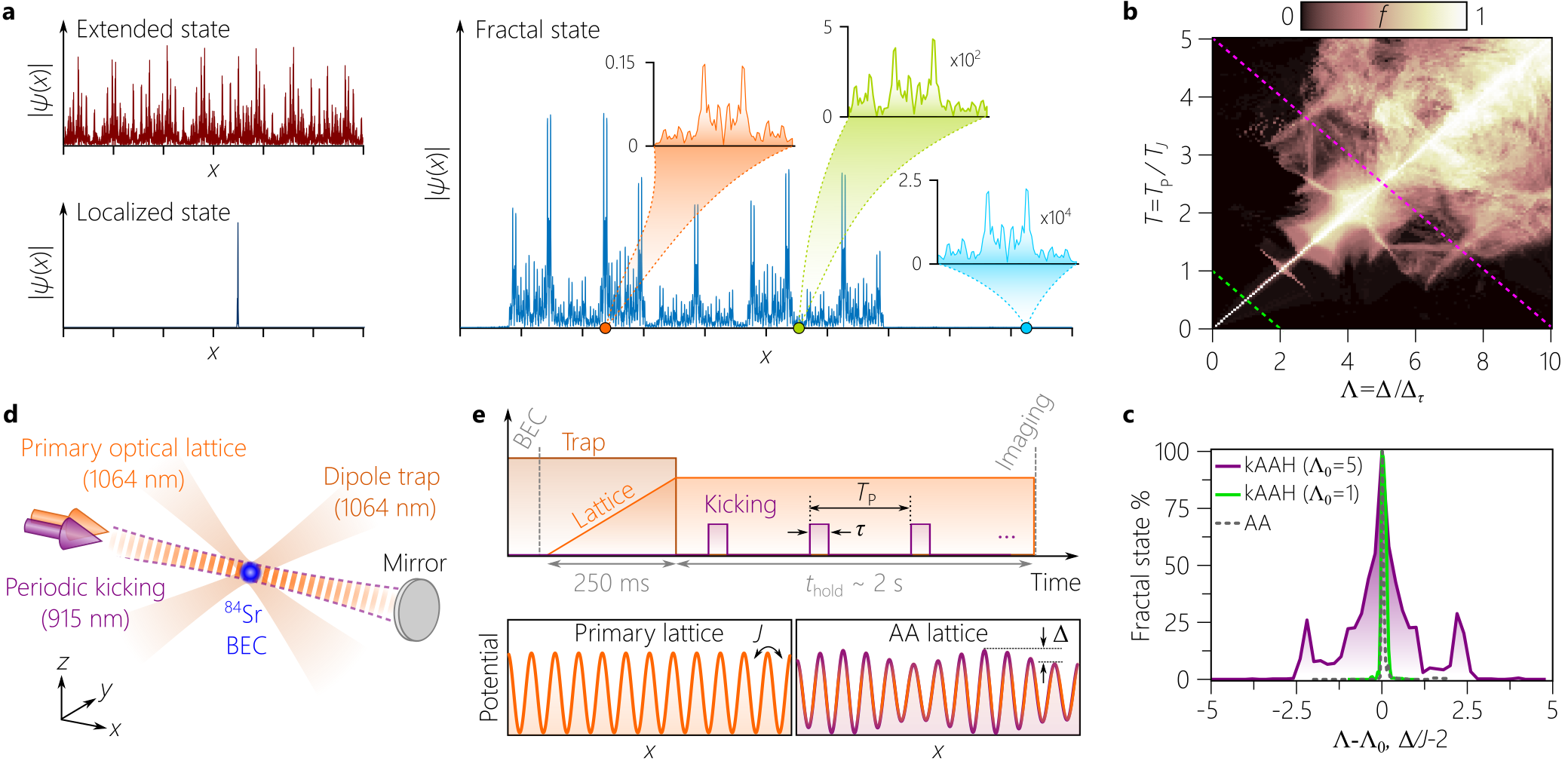}
    \caption{\textbf{Probing multifractality in the kicked Aubry-Andr\'e-Harper model.} 
    \textbf{a}, Examples of extended, localized, and fractal wavefunctions. \textbf{b}, Calculated phase diagram of the kAAH model for  $\alpha=1.162842$. Colorbar shows the percentage of multifractal states, defined operationally by the inequality $0.1 < \xi < 0.9$, versus normalized kick period $T=T_{\mathrm{P}}/T_{\mathrm{J}}$ and kick strength $\Lambda$. \textbf{c}, Percentage of multifractal states versus $\Lambda$, for the static AAH model and the two line cuts of the kAAH phase diagram indicated by dotted lines in \textbf{b}. $\Lambda_0$ corresponds to the point where $\Lambda=T/2$. In the static and rapidly kicked models, fractality occurs only at the phase transition $\Delta/J=2$ and $\Lambda=T/2$, respectively, while at high kick periods an extended multifractal regime emerges. \textbf{d}, Schematic of the experiment. \textbf{e}, Experimental sequence as described in the text. }
    \label{fig:expt} 
\end{figure*}

The time-independent Aubry-Andr\'e-Harper (AAH) Hamiltonian~\cite{aubry1980analyticity,harpermodel} describes particles hopping in a 1D lattice with quasiperiodic pseudo-disorder. This paradigmatic model of quantum transport and quantum Hall phenomena exhibits Hofstadter butterfly energy spectra and hosts critical eigenstates only at a duality-protected localization phase transition~\cite{aubry1980analyticity,harpermodel,hofstadter1976, azbel1979quantum}. If the incommensurate potential is applied instead as brief periodic kicks, the resulting ``occasional quasicrystal'' described by the kicked Aubry-Andr\'e-Harper (kAAH) model epitomizes the interplay between disorder and driving in quantum matter. 
While pioneering experiments in various platforms have explored static and sinusoidally driven AAH models~\cite{roati2008anderson,lucioni2011observation,zilberberg2012,Bordia_Luschen_Schneider_Knap_Bloch_2017,Bloch-SPME,gadwayAA}, the kAAH model has previously been explored only theoretically~\cite{artuso1992phase, artuso1992fractal, leboeuf_phase-space_1990, borgonovi_spectral_1995, Qin_Yin_Chen_2014,cadez_Mondaini_Sacramento_2017}. 

We demonstrate experimentally that the kAAH model exhibits signatures of multifractality across an extended range of parameter space, in agreement both with  numerical calculations and with theoretical expectations for related models~\cite{artuso1992phase, artuso1992fractal, leboeuf_phase-space_1990, borgonovi_spectral_1995}. This letter reports three interconnected advances. The first is the experimental realization of the kAAH model itself. The second is the development of apodized Floquet engineering techniques which use spectrally tailored pulses to extend the accessible parameter range of our quantum simulator into the regime where multifractality is theoretically expected. The third main advance is measurement of the global phase diagram of the kAAH model, which reveals re-entrant delocalization and an extended parameter regime of anomalous transport that we argue is a consequence of multifractal states. 

The experiments begin by loading a Bose-Einstein condensate of 60,000 $^{84}$Sr atoms into the ground band of a 10-$E_r$-deep primary optical lattice, where $E_r=h^2/2m\lambda_P^2$ is the recoil energy, $m$ is the atomic mass, $h$ is Planck's constant, and $\lambda_P=1063.9774(23)$~nm is the primary lattice laser wavelength. To implement the kAAH Hamiltonian, pulsed incommensurate pseudo-disorder is realized by periodically applying a separate overlapped optical lattice of wavelength $\lambda_S=914.4488(17)$~nm. The kAAH Hamiltonian thus realized is given in the tight-binding approximation by
\begin{equation}
	\hat{H}\! =\! -J \sum_{i=1}^L \! \left( \hat{b}^{\dag}_{i} \hat{b}_{i + 1} \! + \! \mathrm{h.c.} \right) + F(t) \Delta \! \sum_{i=1}^L \cos(2 \pi \alpha i + \varphi) \hat{b}^{\dag}_{i} \hat{b}_{i}, \label{kickAAHamiltonian}
\end{equation}
where $J\!=\!0.0192\,E_r\!\approx\!h\!\times\!40$~Hz is the tunneling energy which gives rise to a tunneling time $T_J\!=\!\hbar/J\!=\!(1/2\pi)\!\times\!25$~ms, $\Delta$ is the peak secondary lattice depth, $\alpha=\lambda_P/\lambda_S$ is the wavelength ratio of the two lattices, and $F(t) = \sum_n f_\tau (t-nT_{\mathrm{P}})$ is the waveform of a periodic pulse train composed of finite-width, unit-height pulses with an effective pulse width $\tau=\int f_\tau(t) dt$ and pulse interval $T_{\mathrm{P}}$ (see Supplementary Information Sec.~\ref{suppinfosec1}). As we will discuss, proper selection of the form of $f_\tau(t)$ is crucial for exploring the majority of the phase diagram. We study the phase diagram in a two-dimensional parameter space with dimensionless axes $T=T_{\mathrm{P}}/T_J=T_{\mathrm{P}}J/\hbar$ and $\Lambda=\Delta/\Delta_\tau = \Delta\tau/\hbar$, which characterize the kick period and kick strength respectively. To map localization and multifractality in this unexplored phase diagram, we measure long-time evolution of the density distribution of an initial tightly-confined wavepacket for different $T$ and $\Lambda$ using \emph{in situ} absorption imaging~\cite{roati2008anderson,ketzmerick1997determines,Moessner_fractal_2018}.

Theory predicts a rich phase diagram for this experimentally unexplored model. In the high-frequency limit ($1/T \gg 1$) the model reduces to the static AAH Hamiltonian \cite{Qin_Yin_Chen_2014,cadez_Mondaini_Sacramento_2017}. Much richer behaviors emerge at larger values of $T$ and $\Lambda$ (Figure~\ref{fig:expt}b).  Numerically analyzing the inverse participation ratio (IPR) $\sum_{j=1}^L |\psi_j|^4\!\!\sim\!\!L^{-\xi}$ of single-particle position-space eigenstates $\psi_j$, 
 we find four regimes in the $(T,\Lambda)$ plane, in agreement with previous analyses of the equivalent quantum chaotic kicked Harper model~\cite{artuso1992phase, artuso1992fractal, leboeuf_phase-space_1990, borgonovi_spectral_1995,ketzmerick1999efficient, prosen2001dimer}. The regime with $T\gg \Lambda$ is a completely delocalized phase in which the IPR exponent averaged over all states $\overline{\xi}=1$. This and a duality mapping (Supplementary Information, Sec.~\ref{sec:duality_mapping}) implies a localized phase ($\overline{\xi}=0$) for  $T\ll \Lambda$. 
In between these two limits, for sufficiently large $T$ and $\Lambda$, lies an extended parameter range in which localized and delocalized states co-exist in an intricate nested pattern with a finite fraction of multifractal states (Fig.~\ref{fig:expt}bc)~\cite{ketzmerick1999efficient, prosen2001dimer}. 
At constant $T$ (for example $T=1.5$) the predicted fraction of  localized states can depend non-monotonically on $\Lambda$, suggesting the possibility of anomalous re-entrant delocalized states for which decreasing the disorder strength drives localization. 
In the thermodynamic limit multifractality is expected to persist along the self-dual line $2T=\Lambda$ and in a region of measure zero with a complex shape (Supplementary Information Sec.~\ref{sec:multifractal_fraction})~\cite{ketzmerick1999efficient, prosen2001dimer, artuso1992phase, artuso1992fractal, leboeuf_phase-space_1990, borgonovi_spectral_1995}.
However, the fraction of multifractal states becomes vanishingly small only for extremely large system sizes of $10^7$ sites \cite{ketzmerick1999efficient, prosen2001dimer}. For system sizes relevant to our experiment and even to many conceivable condensed-matter experiments, one finds a non-zero fraction of multifractal states in an extended parameter range away from the duality line. This is consistent with earlier studies which had suggested that an extended multifractal regime persists in the thermodynamic limit~\cite{borgonovi_spectral_1995}. 
The existence and form of this multifractal regime are crucial for understanding our experimental measurements of the kAAH phase diagram: in particular, theory predicts that anomalous transport and re-entrant localization/delocalization transitions should emerge when $T, \Lambda/2 \gtrsim 1$. 

\begin{figure*}[t!]
\includegraphics[scale=1]{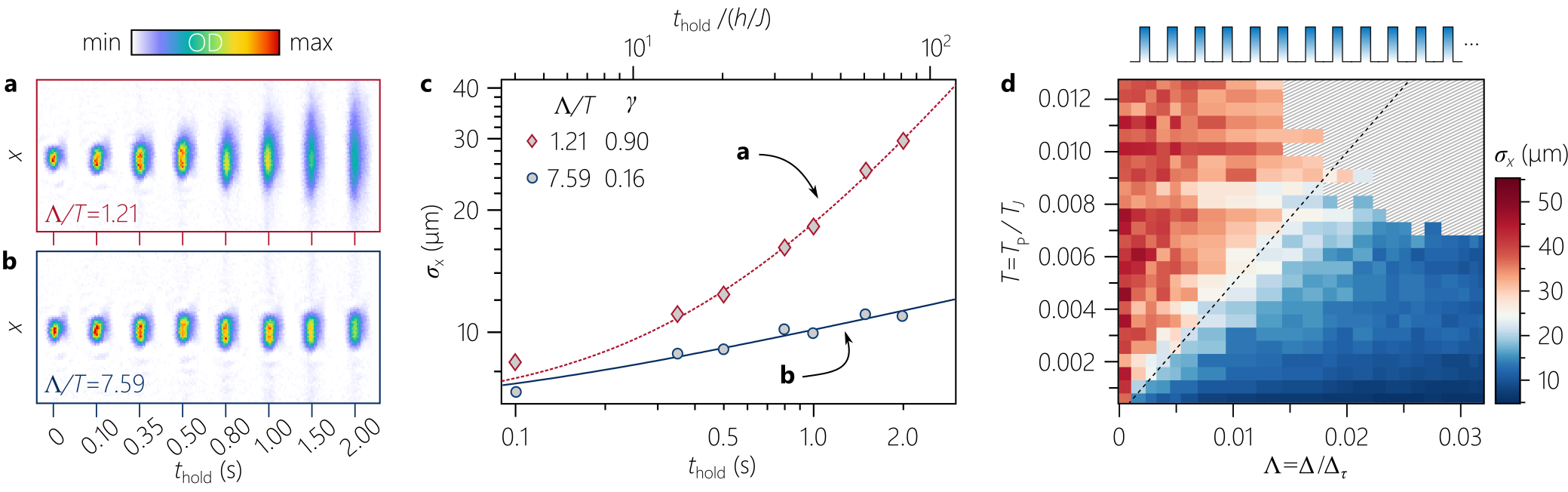}
    \caption{\textbf{Realizing the kicked Aubry-Andr\'e-Harper model in the high-frequency regime.}
    \textbf{a},\textbf{b}, Time sequence of density profiles in the kicked lattice for $\Lambda/T$ below (\textbf{a}) or above (\textbf{b}) the localization transition. \textbf{c}, Fitting expansion curves to a form which captures both short and long-time behavior at various $(T,\Lambda)$ allows for measurement of the exponent $\gamma$. The extracted widths were fit to $\sigma_x(t) \! = \! \sigma_0 (1\!+\!t/t_0)^\gamma$, the solution to a generalized diffusion equation 
    (see Supplementary Information Sec.~\ref{SI-subsec:transport-exponents}). \textbf{d}, Measured localization phase diagram of the kAAH model for small $T$ and $\Lambda$, using a simple rectangular form for the pulse shape $f_\tau(t)$ with $\tau=1$~$\mu$s. Colormap depicts fitted width of the density distribution $\sigma_{x}$ as a function of $\Lambda$ and $T$ at $t_{\mathrm{hold}}\!=\!2~\mathrm{s}$. Dashed line indicates the time-averaged static AA transition at $T\!=\!\Lambda/2$. The center point (white) of the colormap is set to the $\sigma_x$ observed at the same hold time when the expansion exponent is in the center of its transition from localized to delocalized values. Cross-hatched pixels indicate data which failed cuts of the fitting procedure (see Supplementary Information Sec.~\ref{SI-subsec:data-cuts}) due to heating via interband transitions. Without mitigative measures, such heating prevents exploration of the phase diagram much beyond the region shown here; see Fig.~\ref{fig:apodize} for details on characterization and suppression of this effect. }
    \label{fig:2DPhaseDiagram}
\end{figure*}

Figure~\ref{fig:2DPhaseDiagram} encompasses the first main result of our work: experimental realization of the kAAH model.  Figures~\ref{fig:2DPhaseDiagram}a and \ref{fig:2DPhaseDiagram}b show measured density evolution at two points in the phase diagram, one localized and one delocalized. The late-time width clearly distinguishes the two phases. The exponent $\gamma$ characterizing the time evolution of the width $\sigma_x$ of the density distribution ($\sigma_x\propto t^\gamma$ for large $t$) provides another diagnostic, with $\gamma$ near 1 indicating delocalization and $\gamma$ near zero indicating localization~(Fig.~\ref{fig:2DPhaseDiagram}c).  Figure~\ref{fig:2DPhaseDiagram}d demonstrates the first experimental exploration of a swath of the kicked AAH phase diagram, obtained by measuring late-time width of the atomic density distribution for a range of values of $T$ and $\Lambda$ both $\ll 1$, using a simple rectangular form for the kick pulse shape $f_\tau(t)$. In accordance with expectations for this high-frequency regime, 
we observe a period-dependent localization transition near the Aubry-Andr\'e critical point at $\Lambda=2T$. However, our quantum simulator using a rectangular $f_\tau(t)$ breaks down at values of $T$ and $\Lambda$ a hundred times smaller than those where the multifractal regime is predicted to emerge, due to higher-band excitations which invalidate the single-band approximation implicit in the kAAH model. This breakdown is visible in the cross-hatched data points near the upper-right of Fig.~\ref{fig:2DPhaseDiagram}d, which are ($T$,$\Lambda$) pairs where the density distribution failed experimental cuts (see Supplementary Information Sec.~\ref{SI-subsec:data-cuts}) on signal-to-noise ratio and center position, indicating severe heating. Such interband heating prevents experimental access to the rich multifractal regime of the kAAH phase diagram; however, as we show next, this challenge can be overcome.

\begin{figure*}
\includegraphics[scale=1]{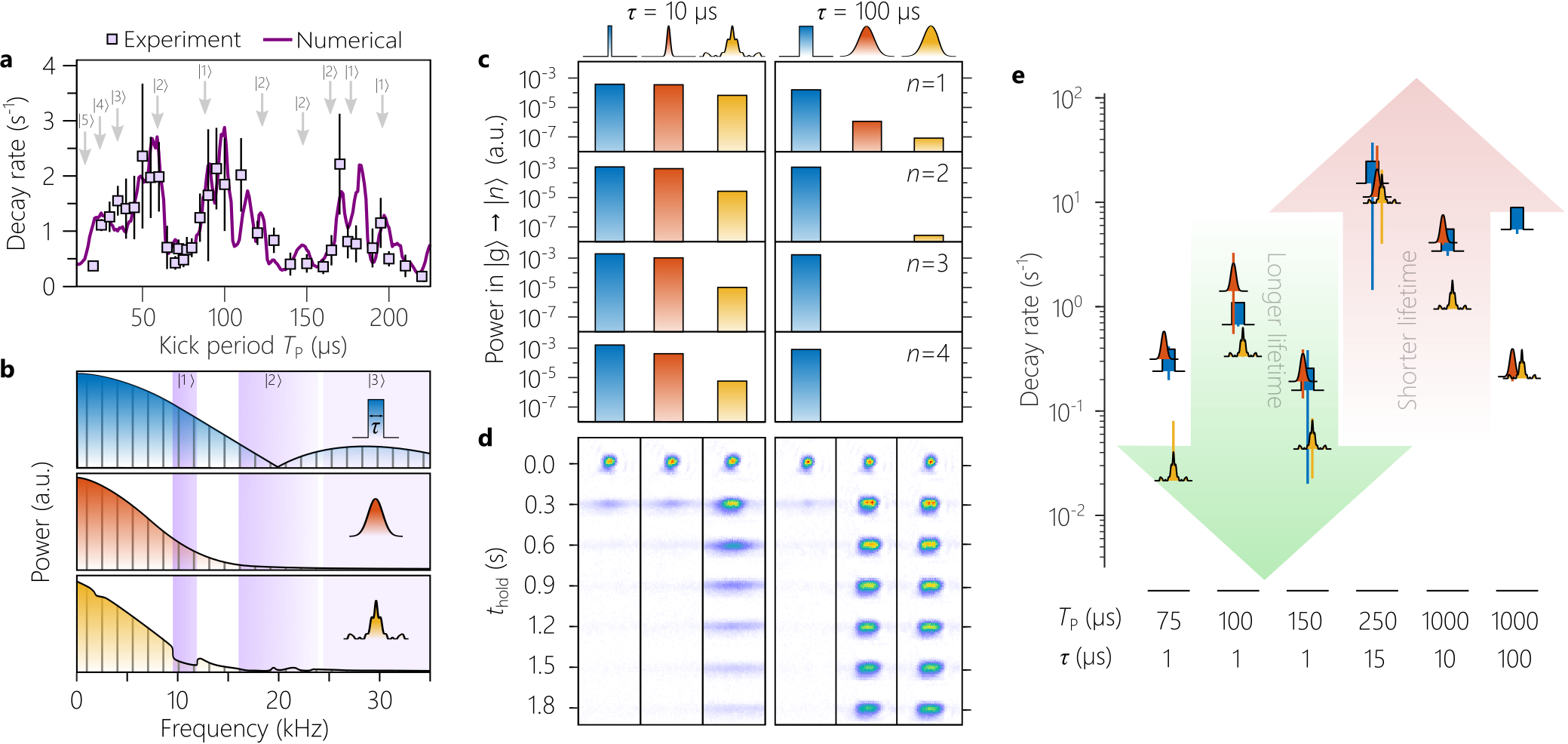}
\caption{\label{fig:apodize} 
\textbf{Apodized Floquet engineering.}
\textbf{a}, Decay rate of atoms kicked by square pulses with a pulse duration of $\tau=1~\mathrm{\upmu s}$ for $t_\mathrm{hold}=2~\mathrm{s}$ as a function of kick period $T_\mathrm{P}$. Solid curve is the result of a numerical calculation (see Supplementary Information Section~\ref{sec:bandcalc}). Arrows indicate dominant transitions to excited bands $|n=1,2,…\rangle$ (see also Extended Data Figure \ref{Fig:DecayTimeDomain}).
\textbf{b}, Form of the power spectrum of square (top), Gaussian (middle) and filtered (bottom) pulses. Shaded areas represent interband transitions. 
\textbf{c}, Net power in frequency ranges corresponding to interband transitions, for square, gaussian, and filtered pulses of two pulse widths, each with period $T = \SI{1}{\milli\second}$. For longer pulses, the gaussian pulse already has little power in the interband transition frequencies, and so filtering has little additional effect. \textbf{d}, Measured density profile at various times for each pulse shape. \textbf{e}, Measured decay rates from the ground band for different pulse shapes at various values of $T_\mathrm{P}$ and $\tau$.}
\end{figure*}

To extend the kAAH quantum simulator to the predicted multifractal regime of the phase diagram with $T, \Lambda/2 \gtrsim 1$ 
we developed a spectrally tailored (``apodized'') Floquet engineering technique which is the second main result of this work and which may prove useful in other contexts. Apodization is a filtering technique common in areas from optics to RF engineering; here we perform an analogous procedure by changing the shape of the pulse function $f_\tau(t)$, thereby modifying the spectrum of the kicking waveform. The goal is the elimination of population transfer to excited bands which breaks the single-band assumption of the kAAH model. Figure~\ref{fig:apodize}a demonstrates the origin of this problem: values of $T$ higher than a few tens of $\mu$s give rise to rapid decay of ground-band atoms. Peaks in the decay rate match numerically-calculated rates of interband transfer, suggesting that the root cause of the heating is the overlap of the spectrum of the kicking waveform with interband transition frequencies (see also Extended Data Figures~2~and~3). To eliminate interband heating, we consider two approaches: (1) conventional apodization with a Gaussian window function and (2) spectral filtering of the pulse train to specifically remove interband transition frequencies. In both cases, the time-domain pulse shape $f_\tau(t)$ is renormalized to give equivalent pulse areas. Figures~\ref{fig:apodize}b,c show the power spectrum of different pulse shapes (square, Gaussian, and spectrally filtered), superimposed (b) and integrated (c) over the frequency ranges of interband transitions, indicating that pulse shaping can address the problem of interband heating. Figure~\ref{fig:apodize}d shows experimental results for density evolution with the three different pulse shapes for different pulse lengths $\tau$, demonstrating the significantly better performance of shaped pulses, and Fig.~\ref{fig:apodize}e shows the measured decay rate for different pulse shapes, kick periods, and repetition rates. Longer lifetimes are  consistently achieved by spectrally engineering the kicking pulse shape. For the kAAH model, apodized Floquet engineering is absolutely required for accessing the nontrivial regions of the phase diagram at large $T$ and $\Lambda$ with quantum gas experiments. 

\begin{figure*}[ht]
\includegraphics[scale=1]{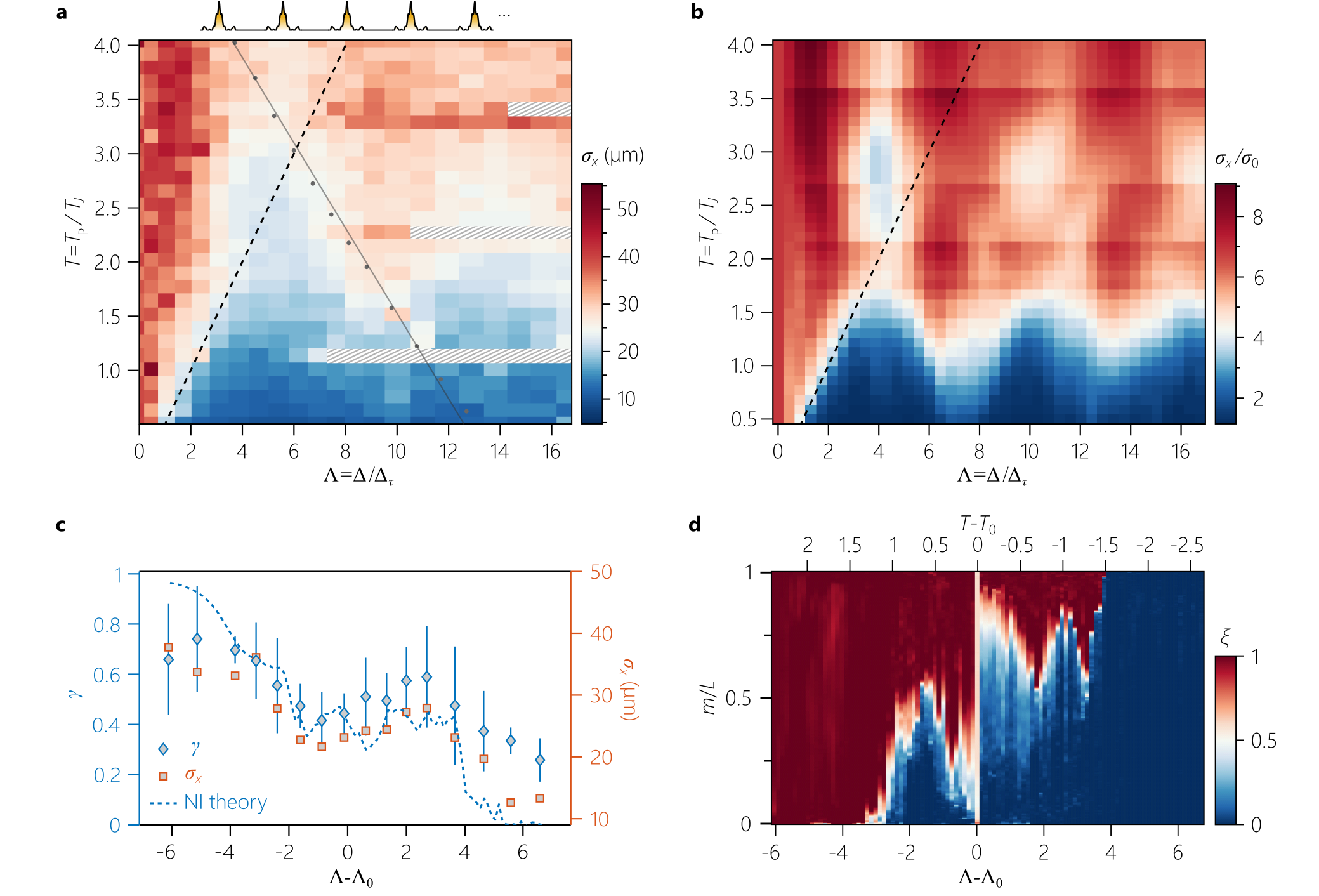} 
    \caption{\label{fig:largeTPhaseMap} \textbf{Experimental signatures of anomalous localization and multifractality in the kAAH model.}
    \textbf{a}, Measured phase diagram of the kAAH model for large $T$ and $\Lambda$ using apodized kicking waveforms. Colormap shows fitted width of density distribution  $\sigma_{x}$ for $t_{\mathrm{hold}}=2~\mathrm{s}$. Dashed line indicates the time-averaged static AAH transition at $T=\Lambda/2$. The colormap center (white) is chosen at this transition point as in Fig.~\ref{fig:2DPhaseDiagram}d. Cross-hatched pixels indicate data which failed cuts (see Supplementary Information Sec.~\ref{SI-subsec:data-cuts}); these appear to be caused by resonant excitation of transverse modes. 
    \textbf{b}, Numerically calculated phase diagram of the kAAH model as a function of $T$ and $\Lambda$ for $\alpha=1.162842$. Colormap shows expansion fraction of an initial Gaussian wave packet with $\sigma_0=1.2$ lattice sites at time $t=10.7~T_J$. Note that while the numerically calculated kAAH phase map shown in Fig.~\ref{fig:largeTPhaseMap}b is not a direct simulation of the experiment, it does assume an evolution time such that the maximal expansion ratio $\sigma_x/\sigma_0$ within the range of the phase diagram is approximately the same in both Fig.~\ref{fig:largeTPhaseMap}a and Fig.~\ref{fig:largeTPhaseMap}b. This approach allows for a fair comparison between experiment and numerical simulation in the cloud width, which is a non-universal quantity. Discrepancies between the experimental and theoretical phase diagrams may also arise from interatomic interactions, finite pulse duration, and a weak axial trapping potential, all of which are present in our experiment but absent in the theory.
    \textbf{c}, Expansion exponent $\gamma$ versus $\Lambda-\Lambda_0$ (diamonds), for parameter values indicated by points in panel \textbf{a}, extracted by fitting the late-time width evolution  to $\sigma_x(t) \propto t^\gamma$. Error bars show 95\% confidence intervals. Measured $\sigma_x$ is also plotted (squares), indicating that this single measurement tracks well with fits of the full time series. \textbf{d}, Scaling exponent $\xi$ of inverse participation ratio for single-particle states as a function of disorder strength $\Lambda$ for the same parameter range. $m/L$ denotes the normalized index of eigenstates.}
\end{figure*}

Figure ~\ref{fig:largeTPhaseMap}a shows phase map data from such an apodized quantum simulator including the predicted multifractal regime.
While the time-averaged Aubry-Andr\'e phase transition at $T=\Lambda/2$ is visible for small $T$, areas of anomalous localization and delocalization appear at larger $T$ in the regime where multifractality is expected. As the disorder strength is increased for $T>1.5$, the system appears to localize, then exhibit re-entrant disorder-driven delocalization, and finally localize again at strong disorder. Both these features agree qualitatively with numerical phase maps of the kAAH model shown in Fig.~\ref{fig:largeTPhaseMap}b. 
To further investigate this newly realized state of driven matter, we measured the full expansion dynamics at selected points on the phase diagram. The transport scaling exponent $\gamma$ serves as a signature of localized, delocalized, and multifractal states in the kAAH model (see Supplementary Information, Sec.~\ref{sec:dynamics})~\cite{artuso1992fractal, ketzmerick1997determines, Moessner_fractal_2018}. 
As shown in Fig.~\ref{fig:largeTPhaseMap}c, across a broad swath of the phase diagram the transport exponent is found to lie between 0.4 and 0.6, which we attribute to the presence of multifractal states. Our experimental observations accord with theoretical predictions also plotted in Fig.~\ref{fig:largeTPhaseMap}c.  Here we note two important subtleties.  First, states in the multifractal regime have mixed character (Fig. \ref{fig:largeTPhaseMap}d), and for any non-zero fraction of delocalized states transport at the longest time scales is expected to be ballistic. However, for experimentally relevant system sizes the presence of multifractal states leads to sub-ballistic transport persisting up to time scales much larger than those explored experimentally (see Supplementary Information Sec.~\ref{sec:multifractal_fraction}). 
Second,  we note that continuously varying transport exponents can also arise in disordered systems due to rare regions in which locally the system resembles a localized state~\cite{agarwal2015anomalous}. However, in a quasiperiodic system such as ours, such rare regions do not exist~\cite{gopalakrishnan2020dynamics}.

These results represent the first glimpse of the rich phase diagram of a prototypical model of localization in driven quantum matter.
The simple but unexpectedly powerful apodization techniques we demonstrate could be applied in a broad range of contexts, and may enhance the range of quantum simulators in any situation where a kicked system emulates a single band model \cite{liu2022anomalous}. Our results raise several fascinating questions for future investigation. Are the multifractal states stable against the introduction of interactions \cite{sarkar2021signatures,gadwaygenAA,zhang2022dynamical} or the presence of gapless propagating modes such as phonons? Can the duality of the phase map be directly imaged? 
In the low-frequency regime the Floquet Hamiltonian is generically long-ranged, and the appearance of a fractal phase recalls predictions of such phases in Hamiltonians with long-range hopping~\cite{deng2019one}. This raises the possibility of engineering other long-range Hamiltonians using the techniques of apodized low-frequency Floquet engineering presented here, potentially allowing the study of phenomena unattainable with local Hamiltonians such as robust error-correcting quantum codes~\cite{hamma2009toric}. A final intriguing possibility for future work is the realization of related fractal states in Floquet-engineered solids or solid-state heterostructures~\cite{basov2017towards}.


\section*{Methods}

\textbf{Preparing the initial state}

The experiments begin by preparing a Bose-Einstein condensate (BEC) of $^{84}$Sr via evaporation in a crossed dipole trap. The BEC is adiabatically loaded  over 250ms into a primary optical lattice. The lattice is formed by retro-reflecting a small portion of a 1064~nm beam, thereby creating a lattice potential superimposed on a single-beam dipole trap. The non-reflected portion of the beam produces the majority of the transverse confinement with a transverse trapping frequency near 60 Hz.  Dynamics are initiated by removing the axial confinement due to the initial dipole trap.

\textbf{Implementing the kAAH Hamiltonian}

The kicking lattice formed by $\lambda_\mathrm{S} =\SI{915}{\nano\meter}$ light is superimposed onto the primary lattice in a periodic pulse train. The pulse intensity, pulse width, and pulse waveform are controlled by tailoring the radiofrequency (RF) waveform sent to the acousto-optic modulator used to modulate the 915nm light intensity. Lattice depths are calibrated by Kapitza-Dirac diffraction. The depth of the primary lattice is stabilized to 10 $E_{R}$ unless otherwise specified, while the depth of the kicking lattice  is monitored by a photodiode and extracted after the experiment. The $\SI{1064}{\nano\meter}$ and $\SI{915}{\nano\meter}$ standing waves share a retro-reflecting mirror, thereby passively stabilizing the relative phase of the two lattices. 

\textbf{Generating apodized waveforms}

To obtain driving waveforms which do not drive excitation to higher bands, unwanted frequency components are filtered out of initial trial waveforms (either Gaussian or square pulses) by means of repeated application of appropriate digital bandpass filters, setting negative amplitudes to zero for physical realizability. The resulting waveform is used as an envelope for the RF signal sent to the acousto-optic modulator which controls the power of the kicking lattice beam. For large $T$ values ($T_\mathrm{P}>\SI{1}{\milli\second}$), simply using a Gaussian waveform is equally effective, as shown in Fig. \ref{fig:apodize}.


\textbf{Data Processing}
The density distribution after time evolution of up to 2 seconds in the kicked lattice is obtained by analyzing \emph{in situ} absorption images. The width of the distribution is extracted from a Gaussian fit to the transversely-integrated density profile. Data was omitted (hashed points in Figs.~\ref{fig:2DPhaseDiagram}a and \ref{fig:largeTPhaseMap}a) for which the center position of this fit differs significantly from the mean, and for which the signal-to-noise ratio was less than 40\% of the mean. Such discarded images are expected to correspond to experimental factors which compromised that particular measurement, or to experimental realizations of the system which produced density profiles so sparse as to not be reliably fitted and analyzed (often due to atom loss associted with significant excitation to higher bands). See Supplementary Information Sec.~\ref{SI-sec:data-analysis} for quantitative details of this procedure.

\begin{acknowledgments}
We thank Laurent Sanchez-Palencia, Dan Arovas, and Andr\'e Eckardt for helpful conversations. We acknowledge support from the Air Force Office of Scientific Research (FA9550-20-1-0240), the Army Research Office (MURI W911NF1710323), and the Eddleman Center for Quantum Innovation, and from the NSF QLCI program through grant number OMA-2016245. We acknowledge support via the UC Santa Barbara NSF Quantum Foundry funded via the Q-AMASE-i program under Grant DMR1906325. This material is based in part upon work supported by the U.S. Department of Energy, Office of Science, National Quantum Information Science Research Centers, Quantum Science Center. T. Grover is supported by an Alfred P. Sloan Research Fellowship and National Science Foundation under Grant No. DMR-1752417. T.-C. Lu acknowledges support from Perimeter Institute for Theoretical Physics. Research at Perimeter Institute is supported in part by the Government of Canada through the Department of Innovation, Science and Economic Development and by the Province of Ontario through the Ministry of Colleges and Universities.
\end{acknowledgments}

\vspace{.3in}

\onecolumngrid

\setcounter{figure}{0}
\renewcommand{\figurename}{\textbf{Extended Data Fig.}}
\renewcommand{\thefigure}{\textbf{\arabic{figure}}}

\begin{figure}[ht]
\includegraphics[scale=1.0]{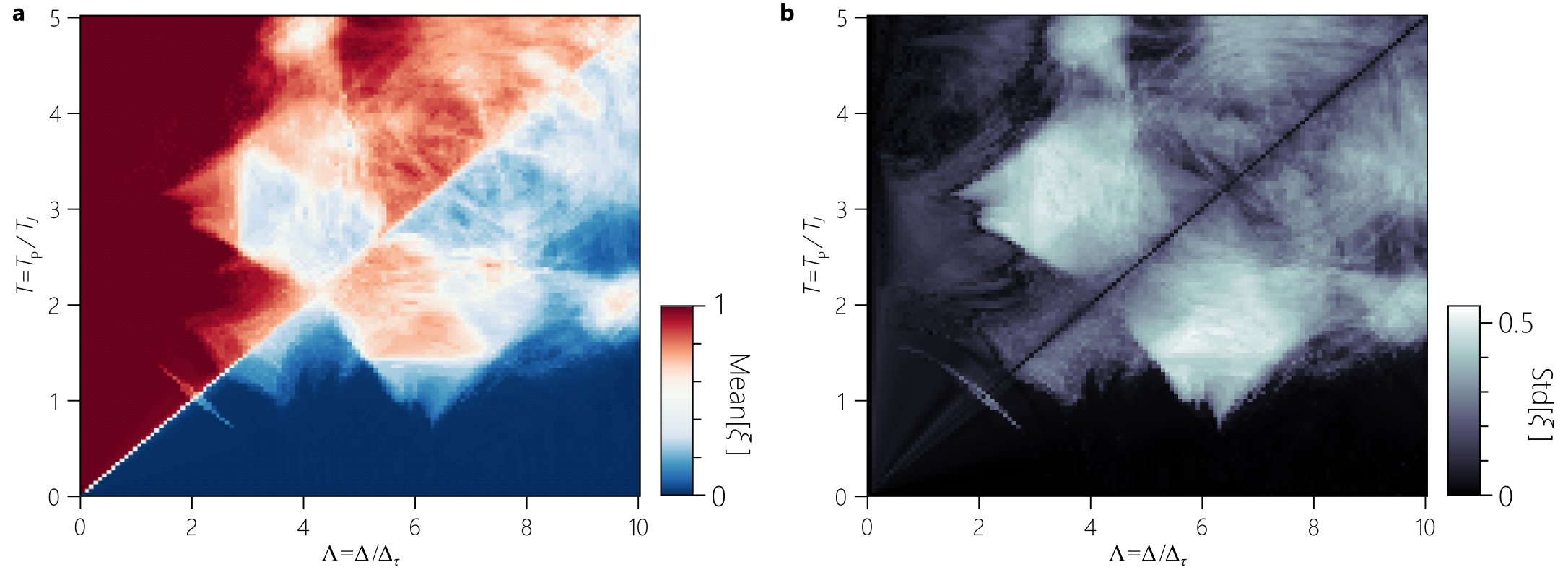}
    \caption{\textbf{Phase diagram of IPR scaling exponents for the kAAH model.} \textbf{a}, Average of the IPR scaling exponents as a function of $T$ and $\Lambda$, obtained by varying the system size from 500 to 1500 in steps of 500 for a lattice period ratio $\alpha=\sqrt{2}/1.21617\approx1064/915$. \textbf{b}, Standard deviation of the IPR scaling exponents. Along the diagonal ($\Lambda=2T$), scaling exponents for all eigenvectors take a single value.   }\label{fig:IPR_Scaling}
\end{figure}

\begin{figure*}
\includegraphics[scale=1]{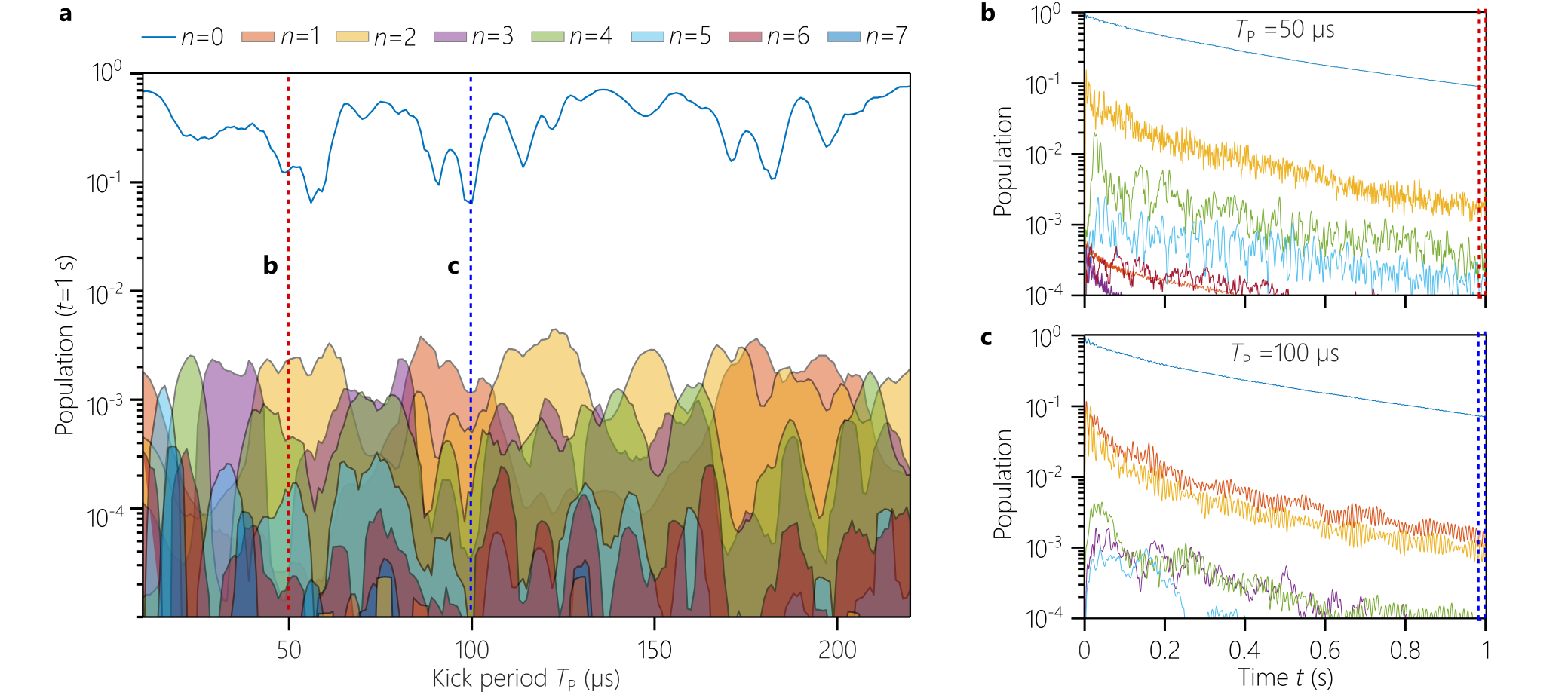}
\caption{\textbf{Multiband time-domain simulations.} Results of numerically simulating band population for 1~s of kicking with $\SI{1}{\micro\second}$ square pulses.
\textbf{a}, Band population distribution as a function of kick period $T_\mathrm{P}$ for the ground and excited bands at $t=1~\mathrm{s}$. 
\textbf{b-c}, Examples of population dynamics for  $T_\mathrm{P}=50~\mathrm{\upmu s}$ (\textbf{b}) and $T_\mathrm{P}=100~\mathrm{\upmu s}$ (\textbf{c}). See Supplementary Information Sec.~\ref{sec:bandcalc} for details of numerical simulations.}
\label{Fig:DecayTimeDomain}
\end{figure*}

\begin{figure*}
\includegraphics[scale=1]{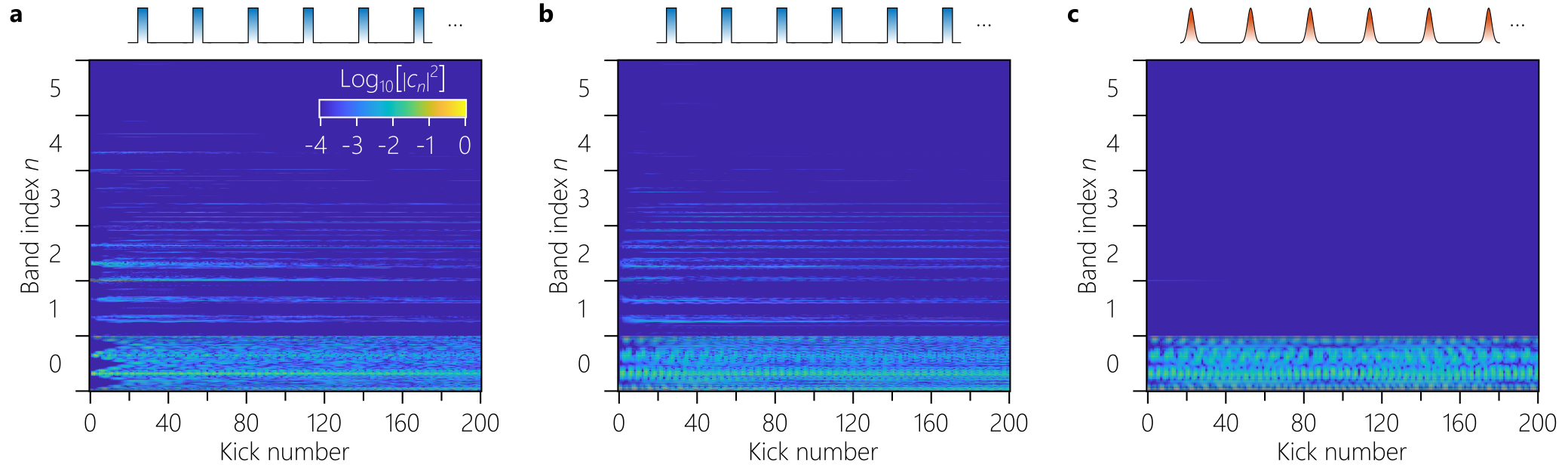}
\caption{\textbf{Spread of the excited eigenmodes with kicking.} \textbf{a},\textbf{b}, Numerical simulations of eigenmode occupation evolution during square-pulse kicking with $\tau=100~\upmu\mathrm{s}$ and $T_\mathrm{P}=1~\mathrm{ms}$ for (\textbf{a})  diabatic and (\textbf{b})  adiabatic loading of the primary optical lattice. Y-axis tick marks correspond to band boundaries.  \textbf{c}, Eigenmode occupation evolution during Gaussian-pulse kicking with equivalent pulse area and kicking period to those of \textbf{a} and \textbf{b}. The primary lattice is adiabatically loaded. While in \textbf{a} and \textbf{b}, the kicking results in excitation to higher bands, in \textbf{c}, excitations are mainly confined to the ground band. See Supplementary Information Sec.~\ref{sec:bandcalc} for details of numerical simulations.}
\label{fig:SpreadOfEigenmodes} 
\end{figure*}

\clearpage

\section*{Supplementary Information}
\setcounter{section}{0}

\setcounter{figure}{0}
\renewcommand{\figurename}{\textbf{Fig.}}
\renewcommand{\thefigure}{S\arabic{figure}}

\setcounter{equation}{0}
\def\theequation{S\arabic{equation}}

\section{Experimentally realizing the \lowercase{k}AAH Hamiltonian}
\label{suppinfosec1}
\subsection{Experimental parameters}
We use ultracold $^{84}$Sr atoms in a bichromatic lattice to construct a quantum simulator of the kicked Aubry-Andr\'e-Harper Hamiltonian. The primary optical lattice is formed by a $\lambda_\mathrm{P}= \SI{1064}{\nano\meter}$ standing wave  (Fig. \ref{fig:expt}) and the secondary kicking optical
lattice formed by $\lambda_\mathrm{S} =\SI{915}{\nano\meter}$ light is superimposed as a periodic pulse train.
Neglecting interatomic interactions, this system is described by the Hamiltonian
\begin{equation}
  \hat{H} = - \frac{\hbar^2}{2m} \frac{\partial^2}{\partial x^2} +  \frac{V_\mathrm{P}}{2} \cos (2 k_\mathrm{P} x) \\ + \frac{V_\mathrm{S} (t)}{2} \cos (2 k_\mathrm{S} x + \varphi)
\label{rshamiltonian}  
\end{equation}
where $k_{P, S} = 2 \pi / \lambda_{P, S}$, $V_{P, S}$ specifies the lattice depths of the primary and the secondary lattice respectively, which are measured in the respective recoil energy of each lattice, $E_{R,i} =\hbar^2 k_i^2/2m$ ($i=P,S)$.
The depth of the primary optical lattice is actively stabilized to 10 $E_{R,P}$ unless otherwise specified, while the depth of the secondary lattice is monitored by a photodiode and measured after each experimental run. Both lattice depths are calibrated by Kapitza-Dirac diffraction. The primary lattice is sufficiently deep, and the initial temperature sufficiently low, that the single-band tight-binding model without a next-nearest-neighbor hopping term is a good approximation~\cite{Holthaus2007PhysRevA.75.063404}. Therefore, the Hamiltonian reduces to the kicked Aubry-Andr\'e-Harper Hamiltonian 
\begin{equation}\nonumber
\hat{H}_{\text{kAAH}} = -J\sum_i ( \hat{b}^{\dag}_{i} \hat{b}_{i + 1} +\text{h.c.} ) +F(t)  \Delta  \sum_i \cos(2 \pi \alpha i + \varphi) \hat{b}^{\dag}_{i} \hat{b}_{i}, 
\end{equation}
where $J$ and $\Delta$ describe the tunneling and disorder strength respectively. Numerical integration of maximally localized Wannier functions is used to extract $J$ and $\Delta$ from lattice parameters~\cite{2009_Modugno}. It is important to note that excitation to higher bands during the experiment will break this single-band approximation and give rise to dynamics not described by the kAAH model; as described in the main text, this motivates careful control of the kicking waveform $F(t)$. The waveform is created by an arbitrary waveform generator with the form: 
\[
F(t) = \sum_n f_\tau (t - nT_{\mathrm{P}})
\]
where $f_\tau(t)$ describes the shape of a single unit-height pulse, $T_{\mathrm{P}}$ is the pulse interval in units of seconds, and $\tau$ is the effective width of a single pulse in units of seconds: $\tau= \int_{-\infty}^\infty f_\tau(t) \, dt$. For example, for square pulses, $\tau$ is simply the pulse duration and for Gaussian pulses with FWHM $w$, $\tau =  \frac{\sqrt{2 \pi}}{2\sqrt{2 \ln 2}} w$.

It is convenient to define dimensionless parameters to describe the kick period and the disorder strength. The kick period is normalized by the tunneling time $T_J$ and the disorder kick strength $\Delta \cdot \tau$ (the disorder depth times pulse width) is normalized by the tunneling energy $J$ times the tunneling time $T_J$: 
\[
T = T_{\mathrm{P}} \frac{J}{\hbar} =\frac{T_{\mathrm{P}}}{T_J}, \quad \Lambda =\frac{\Delta \cdot \tau}{J\cdot T_J}  =\frac{\Delta \tau}{\hbar} = \frac{\Delta}{\Delta_\tau},
\]
where $\Delta_\tau = \hbar/\tau$.
In particular, in the limit of small $\tau$, 
\[
\Delta \sum_n f_\tau (t-nT_{\mathrm{P}}) \rightarrow \sum_n \Delta \tau \delta(t-nT_{\mathrm{P}}) = \sum_n \hbar \Lambda \delta(t-nT_{\mathrm{P}}).
\]

\subsection{Effective Floquet Hamiltonian}

The dynamics of a Floquet system can be analyzed through the one-period Floquet operator:

\begin{equation}\nonumber
    U_F = \mathcal{T} \exp \left[-\frac{i}{\hbar}\int_0^{T_P}  \ H_{\mathrm{kAAH}}(t) \ \mathrm{d}t\right]. 
\end{equation}

For numerical calculations, the finite-width pulse $\Delta \cdot f_\tau(t-nT_{\mathrm{P}})$ in equation~\eqref{kickAAHamiltonian} is replaced by a delta function $\hbar \Lambda \delta(t-nT_{\mathrm{P}})$. For this $\delta$-kicking, the Floquet operator reduces to a product of two operators (taking the time origin at the arrival of the kick):

\begin{equation}
\begin{split}
    U_F &= e^{\frac{i}{\hbar}JT_{\mathrm{P}} \sum_i( -\hat{b}^{\dag}_{i} \hat{b}_{i + 1} +\text{h.c.})}e^{-\frac{i}{\hbar} \Delta \tau \sum_i \cos(2 \pi \alpha i + \varphi) \hat{b}^{\dag}_{i} \hat{b}_{i}} \\
    &= e^{iT \sum_i( -\hat{b}^{\dag}_{i} \hat{b}_{i + 1} +\text{h.c.})}e^{-i \Lambda \sum_i \cos(2 \pi \alpha i + \varphi) \hat{b}^{\dag}_{i} \hat{b}_{i}} = U_{F1} U_{F2} \label{floquetOperator}
\end{split}
\end{equation}

Localization properties can be understood by examining Floquet eigenstates and quasienergies after diagonalizing this Floquet operator (or equivalently the effective Hamiltonian). The localization property of each position-space eigenstate $\psi_j$ can be determined by calculating the system-size dependence of the IPR $\sum_{j=1}^L |\psi_j|^4\!\!\sim\!\!L^{-\xi}$ where $L$ is the system size. Overall localization behavior is then determined by the localization properties of the populated Floquet eigenstates. It is straightforward to extend this analysis for a finite-width pulse or engineered pulse shapes by calculating the Floquet operator numerically. The obtained IPR for kicking with finite-width pulses is qualitatively similar to the $\delta$-kicking case.

For the Floquet operator defined above, one can define the effective Floquet Hamiltonian $H_{\rm{eff}}$ that describes the same dynamics:
\begin{equation}\nonumber
    U_F(T_P,0) = \exp(-\frac{i}{\hbar} H_{\rm{eff}} T_P)
\end{equation}
As discussed in \cite{Qin_Yin_Chen_2014,cadez_Mondaini_Sacramento_2017}, for $T \ll 1$ we can approximately obtain the effective Hamiltonian by using the Baker-Campbell-Hausdorff formula. In general, we can only calculate the effective Hamiltonian numerically. For sufficiently large $T$ values, the effective Hamiltonian has non-zero next-nearest-neighbor hopping terms.  It is thus clear that the kAAH model can be considered as a kind of generalized AAH model~\cite{PhysRevLett.104.070601,PhysRevLett.114.146601,PhysRevB.96.085119,PhysRevB.103.184203}. Generalized AAH models, obtained by modifying the on-site energy or adding long-range hopping terms, can break the Aubry-Andr\'e self duality and give rise to an energy dependent self-duality condition, which in turn gives rise to a single-particle mobility edge. However, as discussed below, this duality is not always broken in generalized AAH models. The kAAH model considered as a Floquet system has another important property. Namely, in addition to the existence of mobility edges, for sufficiently large $T>\pi/2$ the periodic kicks mix different states separated by energy $\hbar \Omega = \hbar (2 \pi)/T_{\mathrm{P}}$ within the first band of the primary lattice, the width of which is $4J$. This effect further enriches the quasienergy spectrum. Namely, the mixing causes multiple mini bands of different localization properties to appear alternatingly, and multiple mobility edges to appear in the Floquet Brillouin zone. At the same time, this coupling mixes Floquet eigenstates with different localization character \cite{Vershinina2017EJP}. Such states with a mixed character are essential in causing delocalization behavior when the initial states are prepared in a localized region. This effect of Floquet mixing also depends on the kick strength $\Lambda$. 

\subsection{Duality mapping in the kAAH model}\label{sec:duality_mapping}
It is well known that the AAH model features a self-duality between position space and momentum space, which pins the energy-independent localization transition at $T = \Lambda/2$. Here, we show that this duality in the AA model is in fact preserved in the $\delta$-kicked AAH model. Given the kAAH model Hamiltonian (Eq.~\eqref{kickAAHamiltonian} of the main text), the corresponding Floquet unitary $U_F$ is given by Eq.~\ref{floquetOperator}. Fixing $\varphi=0$ for convenience, and imposing periodic boundary conditions on the 1d lattice of size $L$, we can perform a Fourier transform for the bosonic operators:
\begin{equation}\nonumber
b_j   = \frac{1}{\sqrt{L}} e^{ -i \pi L j } \sum_k e^{ -i(j+ \pi L) k } \tilde{b}_k,
\end{equation}
where $k=\frac{ 2\pi n}{L}$ for $n=0, 1, 2, \cdots, L-1$. It follows that the hopping term can be written as 
\begin{equation}\nonumber
U_{F1}= \exp [-i T \sum_j   ( - b_j^{\dagger} b_{j+1}  +\text{h.c.}  ) ] =  \exp [-2i T  \sum_k  \cos(k) \tilde{b}_k^{\dagger}\tilde{b}_k ].
\end{equation}
On the other hand, the on-site potential term is 

\begin{equation}\nonumber
	U_{F2}  =  \exp[ -i \Lambda \sum_j    \cos(2\pi \alpha j     )      b_j^{\dagger}b_j ]   = \exp [ -i  \Lambda \sum_{k,q} \{ - V(k,q ) \tilde{b}_k^{\dagger } \tilde{b}_q \} ],
\end{equation}
where $V(k,q) = \frac{ 1 }{2L    }	 \sum_j\left[   e^{  i (k-q +2\pi \alpha) j  } + e^{  i (k-q -2\pi \alpha) j  }    \right]$. Choosing $\alpha =M/L$, where $M$ is an integer, and $M$ and $L$ are coprime and summing over $j$ gives $V(k,q)  = \frac{1}{2} \left[  \delta_{k-q+2\pi \alpha,0}  + \delta_{k-q- 2\pi \alpha,0}      \right]$, and therefore
\begin{equation}\nonumber
	U_{F2}  = \exp \left[-i  \frac{\Lambda}{2} \sum_{  k   } ( - \tilde{b}_k^{\dagger }  \tilde{b}_{ k+ 2\pi \alpha  }   + \text{h.c.} )  \right].
\end{equation}
This operator describes particles hopping in the momentum space via $k\to k\pm  2\pi \alpha =  k\pm  2\pi \frac{M}{L} $.  In other words, in the momentum space lattice labeled by $n =0,1,2,\cdots, L-1$ with $k= 2\pi \frac{n}{L}$, the above hopping corresponds to sending $n \to n  \pm M$. This motivates us to define the new lattice coordinates $\tilde{n}=0,1,2, \dots L-1$ which correspond to the momentum lattice sites $0, M, 2M, \cdots (L-1)M$ mod $L$ so that the hopping becomes nearest-neighboring hopping in $\tilde{n}$ coordinates. Note that such a relabeling requires the choice $\alpha= \frac{M}{L}$ with $M$, $L$ being coprime. For example, choosing $L=5$, $M=2$, the new coordinates $\tilde{n}=0,1,2,3,4$ correspond to the momentum lattice sites $0,2,4,1,3$. On the other hand, when $L$ and $M$ are not coprime, say choosing $L=6$, $M=2$, the new coordinates $\tilde{n}=0,1,2,3,4$ will give the lattice sites $0,2,4,0,2,4$, which thus cannot enumerate all possible momenta.  
 
Finally, defining  the bosonic operators  $\mathrm{b}_{\tilde{n}}$ under the new momentum coordinates $\tilde{n}$ via $\mathrm{b}_{\tilde{n}}=\tilde{b}_k$, one finds 

\begin{equation}\nonumber
		U_{F2}  =  \exp \left[ -i   \Lambda \sum_j    \cos(2\pi \alpha j     )      b_j^{\dagger}b_j  \right] = \exp \left[- i \frac{\Lambda}{2} \sum_{  \tilde{n} =0   }^{L-1}   ( - \mathrm{b}_{\tilde{n}}^{\dagger }  \mathrm{b}_{\tilde{n}+1}   + \text{h.c.} )  \right],
\end{equation}
and 
\begin{equation}\nonumber
U_{F1}= \exp \left[	- i T   \sum_j   ( - b_j^{\dagger} b_{j+1}  +\text{h.c.}  ) \right] = \exp \left[ -2i T \sum_{\tilde{n}=0}^{L-1}   \cos( 2\pi  \alpha \tilde{n}  ) \textrm{b}_{\tilde{n}}^{\dagger}  \textrm{b}_{ \tilde{n} } \right]. 
\end{equation}

This indicates that the hopping term in position space corresponds to the on-site potential term in momentum space, and the on-site potential term in position space corresponds to the hopping term in momentum space. The self-dual point is given by $2T = \Lambda$.

The above derivation shows that the exact duality in a system of finite size $L$ requires $\alpha=\frac{M}{L}$ with $M$, $L$ being coprime. The same self-duality also arises naturally in the equivalent kicked Harper model which will be introduced in the next section \cite{artuso_fractal_1994, prosen2001dimer}. In the experiment, $\alpha$ is fixed at $\lambda_P/\lambda_S=1.163518(5)$, indicating that the exact duality holds for the experimental system size. In practice, a localization transition can be observed when the number of lattice sites is large enough within the actual periodicity of the potential and the system size is not much greater than the spatial period (see e.g. \cite{2009_Modugno} for a discussion of this point).  

\subsection{Equivalence between the kicked Harper Model and the kicked AA model}
Here we show that the model given by Eq.~\eqref{kickAAHamiltonian}, in the $\delta$-kicking limit, is equivalent to the kicked Harper (kH) model \cite{leboeuf_phase-space_1990, artuso_fractal_1994, borgonovi_spectral_1995}, a paradigmatic model of quantum chaos. This relationship was already briefly noted for example in \cite{Qin_Yin_Chen_2014}; we outline an explicit derivation below. 

The kH model is described by the Hamiltonian
\[
\hat H_{\text{kH}} = L \cos (2 \pi \alpha \hat n) + K \cos (\hat \theta) \sum_{j = -\infty}^\infty \delta (t - j).
\]
Here for convenience we set $\hbar = 1$. Operators $\hat n$ and $\hat \theta$ are momentum and angular coordinate operators, respectively, which are canonically conjugate: $[\hat n, \hat \theta] = i$; $\alpha$ is again the incommensurate ratio. The one-period Floquet operator of the kH model is
\[
U_{\text{kH}} = e^{-i L\cos (2 \pi \alpha \hat n)}e^{-i K\cos (\hat \theta)}.
\]
Similar to the static Harper Hamiltonian, the kH model also features the duality along the line $L = K$ since $U_{\text{kH}}$ is formally unchanged under the exchange $K \leftrightarrow L$. 

To demonstrate the desired equivalence we will work in the momentum basis $\hat n \vert n \rangle = n \vert n \rangle$. Let $\hat b_i$ be the boson annihilation operator at site $i$. Within this second-quantization approach, the two terms in the kH model can be mapped to
\begin{align*}
    L \cos (2 \pi \alpha \hat n) &\Rightarrow L \sum_{n\in \mathbb{Z}} \cos (2 \pi \alpha n) \hat b_n^\dagger \hat b_n,\\
    K \cos \hat \theta &\Rightarrow K  \sum_{m,n \in \mathbb{Z}} \langle m \vert \cos \hat \theta \vert n \rangle \hat b^\dagger_n \hat b_m. 
\end{align*}
The matrix element $\langle m \vert \cos \hat \theta \vert n \rangle$ can be evaluated as
\begin{align*}
    \langle m \vert \cos \hat \theta \vert n \rangle &= \int \mathrm{d} \theta \langle m \vert \theta \rangle \langle \theta   \vert n \rangle\cos \hat \theta \\
    &= \frac{1}{2\pi} \int \mathrm{d} \theta \, \frac{e^{i \theta} + e^{-i \theta}}{2} e^{i (m - n) \theta} \\
    &= \frac{1}{2} \left(\delta_{m, n+1} + \delta_{m, n-1}\right), 
\end{align*}
so the $K$ term becomes
\[
K \cos \hat \theta \Rightarrow \frac{1}{2} K \sum_{n \in \mathbb{Z}}\left(\hat b^\dagger_{n+1} \hat b_n + \text{h.c.} \right). 
\]
We thus see that the momentum operator $\hat n$  in the kH model maps to the real-space lattice position in the kAAH model, both of which have integer-valued spectra. Similarly, the angular variable $\hat \theta$ in the kH model maps to the lattice momentum in the kAAH model, both of which are defined modulo $2 \pi$. The Floquet operator becomes
\[
U_{\text{kH}} = e^{-i L \sum_{n} \cos (2 \pi \alpha n) \hat b_n^\dagger \hat b_n} e^{-i \frac{K}{2} \sum_{n}\left(\hat b^\dagger_{n+1} \hat b_n + \text{h.c.} \right)}.
\]
The equivalence between the kH model and the kAAH model as defined in the equation~\eqref{kickAAHamiltonian} of the main text is then established by choosing
\[
L = \Lambda, \quad K = 2 J T. 
\]
The line of duality corresponds to $\Lambda = 2T$ when $J = 1$. 

\section{Numerical Modeling }

\subsection{Calculation of interband excitation}

\label{sec:bandcalc}

Here we present the method for numerical calculation of the rate of (undesirable) interband excitations in our quantum simulator of the kAAH model. Understanding this behavior is critical both because accurate quantum simulation of the kAAH model requires the atoms to remain in the ground band and because interband heating results in atom loss, prohibiting observation of long-time dynamics. We use the standard eigenvector decomposition method to calculate dynamical evolution in the periodically kicked lattice. For square-pulse kicking, we consider two time-independent optical potentials, namely the primary optical lattice and the quasiperiodic (kicked) lattice, given by 
$$ V(x)= 
\begin{cases}
	\frac{V_\mathrm{P}}{2} \cos (2 k_\mathrm{P} x),& \text{for the primary lattice only}\\
	\frac{V_\mathrm{P}}{2} \cos (2 k_\mathrm{P} x)  + \frac{V_\mathrm{S}}{2} \cos (2 k_\mathrm{S} x + \varphi), & \text{for the quasiperiodic (kicked) lattice}.
\end{cases} $$
The time dynamics are obtained by projecting the Schr\"odinger equation at the boundaries between the two Hamiltonians constructed based on the potentials given above and using the time-dependent solution of the Schr\"odinger equation given by  $\Psi(x,t)=\sum_n c_n \psi^{(n)}(x) e^{-iE_nt/\hbar},$ where $\psi^{(n)}(x)$ is the $n^{\mathrm{th}}$ eigenvector with the corresponding eigenvalue $E_n$. 

As a first step to realize multiband simulation of the one-dimensional kAAH Hamiltonian, we discretize the optical potential with 256 lattice sites such that it results in eigenvectors and eigenmodes for the first 12 bands. For this purpose, we set the mesh size along $x$, the optical lattice direction, to be $\mathcal{M}=2\times 12 \times 256 +1=6145$. We include the effect of the  optical lattice beam's Rayleigh range ($\approx 25~\mathrm{mm}$) on the overall optical potential. After constructing the Hamiltonian as a sparse matrix with the specified number of mesh points, we calculate only the smallest $12\times 256 =3072$ eigenvalues and eigenvectors. We estimate that 12 bands and the corresponding number of eigenvectors is sufficient for simulating the dynamics of quasicrystals kicked by square pulses with widths down to $1~\mathrm{\upmu s}$. For Gaussian pulses, the required number of bands can be smaller due to the relative lack of high-frequency spectral weight. 

\begin{figure}[htb]
	\centering
	\includegraphics[scale=1]{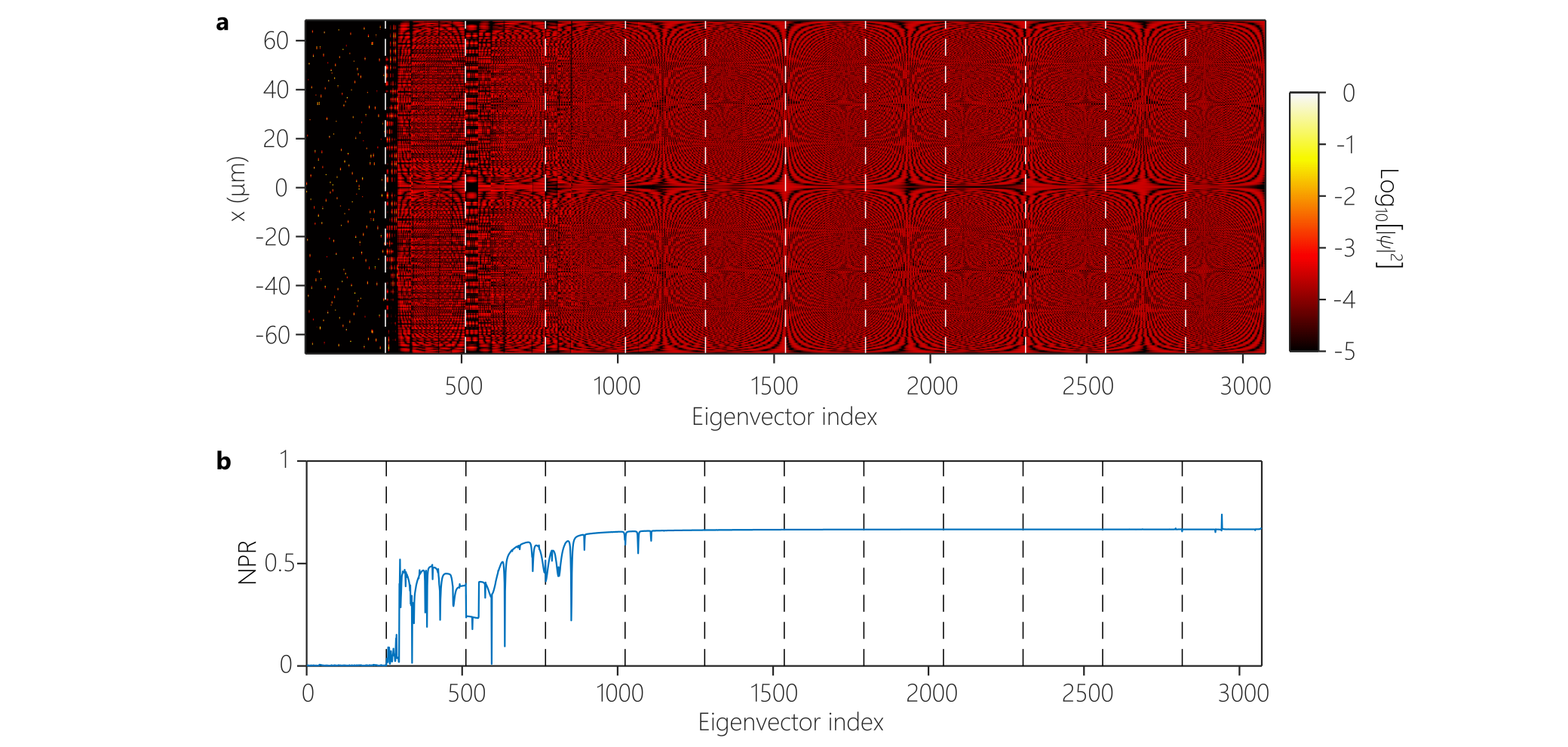} 
	\caption{\textbf{a}, False-color image of the eigenvector amplitudes for the first 12 bands for primary optical lattice depth $s_1=10E_\mathrm{R,P}$ and secondary optical lattice depth $s_2=0.687E_\mathrm{R,S}$  corresponding to $\Delta = 0.277E_\mathrm{R,P}$ and $\Delta/J=14.43$.  \textbf{b}, Corresponding NPR. \textbf{a},\textbf{b}, The dashed lines separate the bands.}
	\label{Fig.S.Eigenvectors}
\end{figure}

We use the normalized participation ratio (NPR) to define the decay rate of the kicked quantum gas. For  the $n^{\mathrm{th}}$ eigenvector, NPR is given by $$\mathcal{NPR}^{(n)} = \left[\mathcal{M}\sum_{m=1}^{\mathcal{M}} |\psi_m^{(n)}|^4 \right]^{-1}.$$ 
Note that we define NPR based on the actual meshing number instead of the number of lattice points. Atoms excited to higher bands tend to rapidly escape from the finite-depth ($\approx\!10E_\mathrm{R,P}$) optical potential; for reference, eigenvalues at the  $6^{\mathrm{th}}$ band have values near or above $20E_\mathrm{R,P}$ and eigenvalues at the $12^{\mathrm{th}}$ band have values around $100E_\mathrm{R,P}$ for moderate kicking strengths. Moreover, eigenvectors in excited bands are spatially much less confined, especially for the kicked lattices, and the tunneling time reduces as the band index increases. As a result, higher-band atoms delocalize rapidly and either leave the trap or become so diffuse as to be difficult to experimentally detect . For these reasons, we assume that the decay rate is approximately proportional to the NPR value of a given eigenvector. Figure~\ref{Fig.S.Eigenvectors} depicts eigenvectors and the associated NPR values during kicking. 

\begin{figure}[htb]
	\centering
	\includegraphics[scale=1]{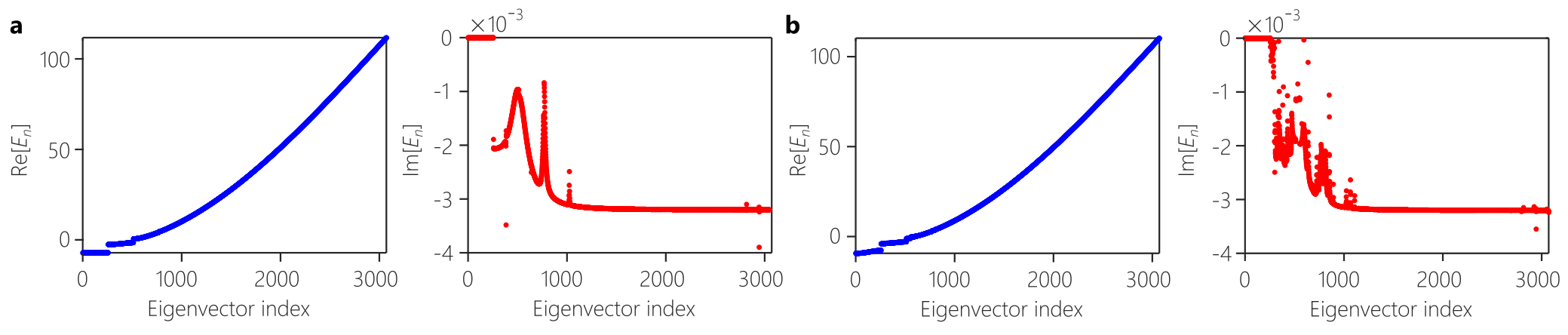} 
	\caption{\textbf{a},\textbf{b}, Modified eigenvalues (in units of $E_\mathrm{R,P}$) for the first 12 bands.  \textbf{a}, Primary optical lattice depth  $s_1=10E_\mathrm{R,P}$ in the absence of  secondary optical lattice  ($s_2=0$). \textbf{b}, Primary optical lattice depth  $s_1=10E_\mathrm{R,P}$ and secondary optical lattice depth $s_2=0.687E_\mathrm{R,S}$  corresponding to $\Delta = 0.277E_\mathrm{R,P}$ and $\Delta/J=14.43$.}
	\label{Fig.S.EigenvaluesMod}
\end{figure}

To simulate decay dynamics due to $1~\upmu s$ square pulses (Fig.\ref{fig:apodize}a of the main text), we  modify the eigenvalues such that $E_n\rightarrow E_n-i\kappa \mathrm{K}(\mathcal{NPR}^{(n)})$. Here, $\kappa$ is an effective extinction coefficient that results in decay of the population in the time domain simulations, and $\mathrm{K}(x)=\left[x-\mathrm{min}(x)\right]/\mathrm{max}\left[x-\mathrm{min}(x)\right]$. This leaves only a single parameter to be fitted to the experimental results; we find the optimum value of the extinction coefficient to be $\kappa=3\times 10^{-3}$. Figure~\ref{Fig.S.EigenvaluesMod} depicts the modified eigenvalues for the primary optical lattice only and the kicked optical lattice. Note that, after the first 4 bands, the NPR value saturates for both lattices. 

To obtain the final decay coefficients (shown as a purple line in Fig.~\ref{fig:apodize}a), we fit the population remaining in the ground band to a simple exponential decay function ($\propto e^{-\alpha t}$) without an offset using a non-linear least squares method. To make the fitting procedure more robust and eliminate the non-steady effects of periodic kicking, we  disregard the initial 10 percent of the total kicking duration.

\subsection{Calculation of transport dynamics}\label{sec:SI_simulation_method}

Here we present the method for numerical calculation of transport in the kAAH model that gives the theoretical predictions presented in Fig.~\ref{fig:largeTPhaseMap}b and Fig.~\ref{fig:largeTPhaseMap}c. To  model the experimental setup, where a Bose-Einstein condensate of $N$ bosons is initially localized in a spatial subregion, theoretically we  consider the initial state $ | \psi_0 \rangle$ created by filling $N$ bosons into a vacuum, namely, $ |\psi_0 \rangle  =   \frac{1}{ \sqrt{ N!  }  }   (b^{\dagger})^N | 0  \rangle $. Here $b^{\dagger}$ is a boson creation operator defined as $b^{\dagger}= \sum_{i=1}^L \phi_i b^{\dagger}_i $ so that a single particle created using $b^{\dagger}$ has a real-space wave function given by $\phi_i$, where $\sum_i |\phi_i|^2=1$ and the boson satisfies the conventional commutation relation, i.e.  $[b,b^{\dagger}]=1$. To simulate the evolution of the real-space density profile of bosons under the kAAH dynamics, we compute the number operator $\langle \psi(t) |b_i^{\dagger} b_i| \psi(t) \rangle $, where $| \psi(t) \rangle  $ is the time-evolved state, i.e. $| \psi(t) \rangle = U(t)|\psi_0\rangle$, and $U(t)$ is the time evolution operator. The quantity $\langle \psi(t) |b_i^{\dagger} b_i| \psi(t) \rangle $ can be written as $ \langle b_i^{\dagger }(t) b_i (t)  \rangle  =\langle \psi_0 |b_i^{\dagger} (t)b_i (t)|  \psi_0 \rangle$, where $b^{\dagger}_i(t)= U^{\dagger}(t)b^{\dagger}_i U(t)$ is the time-evolved operator in the Heisenberg picture. Since the theory neglects the effects of interactions, the many-body unitary evolution reduces to a single-particle evolution in real space, i.e. $U^{\dagger}(t)b_i^{\dagger} U(t)= \sum_j u_{ij} b^{\dagger}_j$, where $u$ is a $L\times L$ unitary matrix derived from the kAAH model. It follows that the real space number distribution of bosons is $ \langle  b_i^{\dagger}(t) b_i(t)  \rangle = \sum_{ jk }  u_{ij}  u^{*}_{ik}  \langle b_j^{\dagger} b_k   \rangle  $ with $ \langle b_j^{\dagger} b_k   \rangle   =  \langle\psi_0 |  b_j^{\dagger} b_k  |  \psi_0 \rangle  =N \phi_j\phi_k^*$, where the last equality is derived using the commutation relation  $	[b_j,  b^{\dagger}  ]=  \phi_j$. Therefore one derives the time-evolved boson number distribution in real space:
\begin{equation}
\langle  b_i^{\dagger}(t) b_i(t)  \rangle  = N \sum_{jk} u_{ij}u_{ik}^*  \phi_j\phi_k^* =  N \lvert \sum_{j} u_{ij} \phi_j \rvert^2.
\end{equation}
Using this real-space distribution, one can compute the variance $\sigma^2(t)$ for any $t$. For comparision with experimental data we choose  $\phi_i \sim  e^{ \frac{-( i-i_0 )^2}{ 4\sigma_0^2  }   }$ so that the initial density distribution $|\phi_i|^2$ is a Gaussian centered at the $i_0$-th site with standard deviation $\sigma_0$.

\subsection{Comparison of measured and calculated transport exponents}
Here we discuss and evaluate a few of the most important reasons for discrepancies between measured and theoretically calculated transport exponents: slight experimental imperfections in the kick uniformity, effects of coherence in the initial state, and interparticle interactions. 

\subsubsection{Random disorder in $\Lambda$}
\begin{figure}[ht]
\includegraphics[scale=0.35]{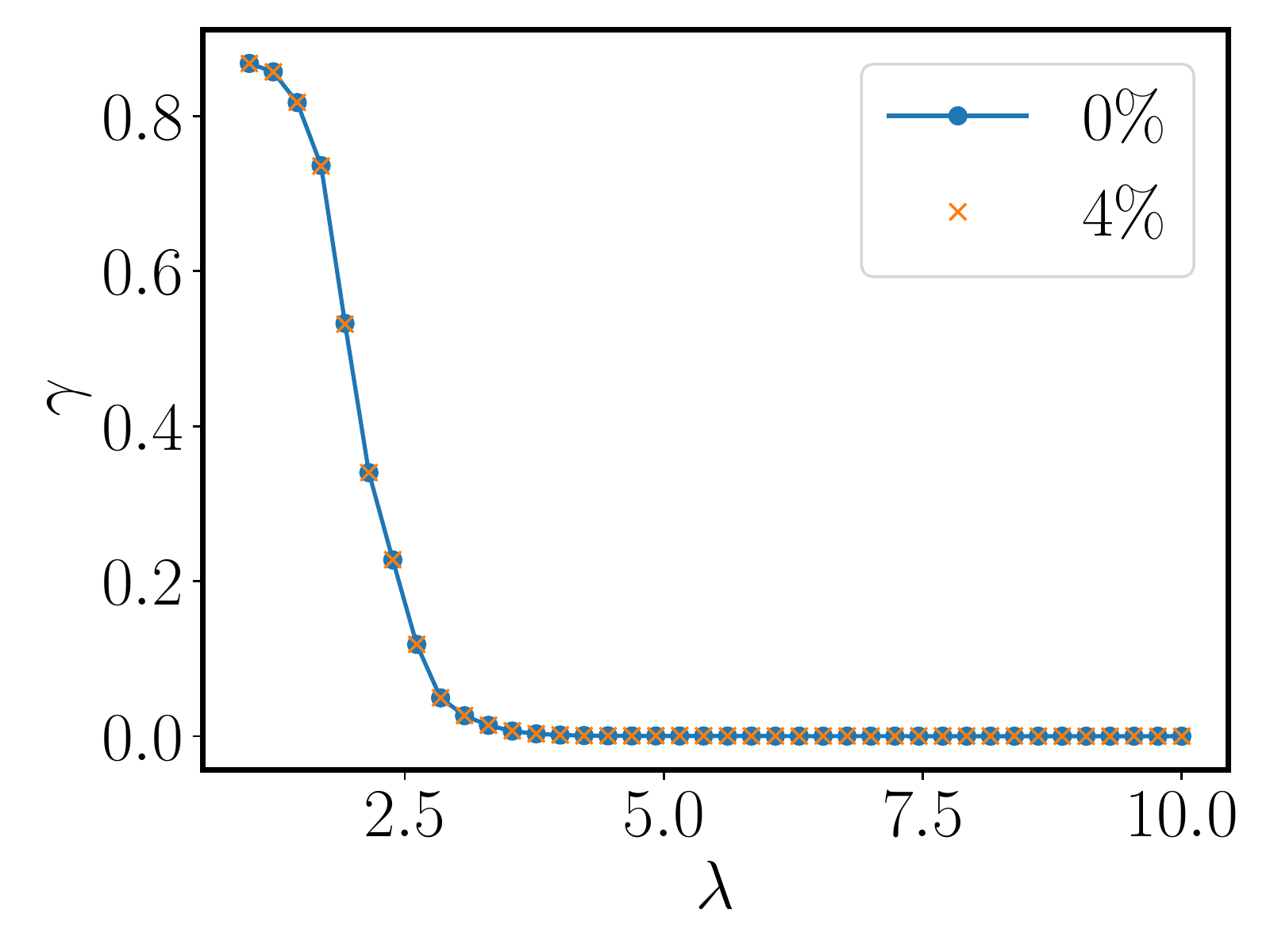}
    \caption{Numerically computed transport exponents with and without random variations in $\Lambda$.  }\label{fig:appendix_random_lambda}    
\end{figure}

In a typical sequence of experiments performed for a nominally constant $\Lambda$, we measure about $4\%$ variation in the actual kick strength due to thermal effects and laser power fluctuations. We have performed a numerical simulation to check the effect of such  randomness in $\Lambda$ on the transport exponent. Fig.\ref{fig:appendix_random_lambda} shows the numerical result both with and without such randomness. We find that random variations in $\Lambda$ at the experimentally relevant level do not cause any significant difference in the transport exponent. 

\subsubsection{Finite-time effect in coherent initial states}

\begin{figure}[ht]
\includegraphics[scale=0.25]{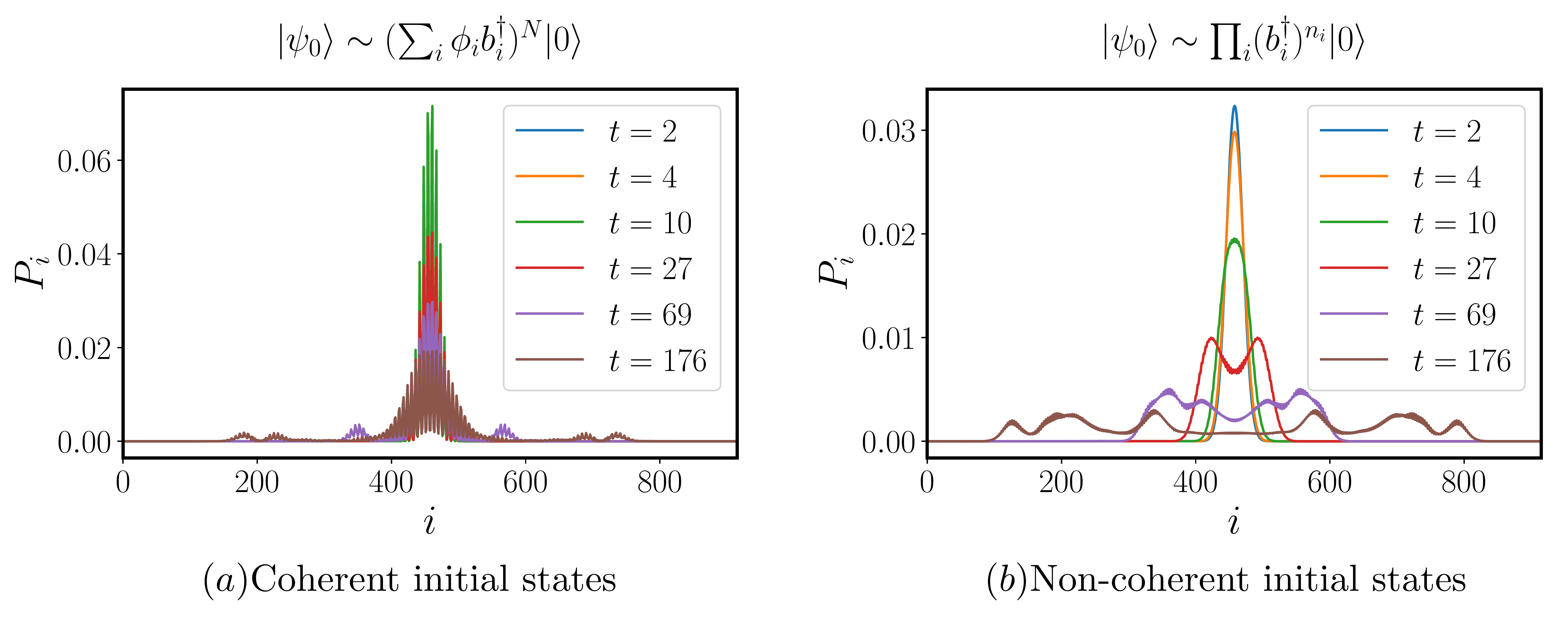}
	\caption{Comparison in the spreading of the boson density profile in real space between a coherent initial state (i.e. $|\psi_0 \rangle \sim (\sum_i \phi_i b_i^{\dagger})^{N} |0 \rangle$) and an initial state without spatial coherence (i.e. $|\psi_0 \rangle \sim \prod_i (b_i^{\dagger})^{n_i} |0 \rangle $). 
	}  \label{fig:SI_bec_mott}\end{figure}

Here we provide numerical data showing that the coherence among lattice sites in the initial state described by a Bose-Einstein condensate leads in general to slower transport, and that therefore, ballistic transport can only be observed at a late time. In Fig.\ref{fig:SI_bec_mott}(a), we consider a coherent initial state $|\psi_0 \rangle \sim (\sum_i \phi_i b_i^{\dagger})^{N} |0 \rangle $ by filling $N$ identical bosons of wave function $\phi_i$ into the vacuum, and observe a slower spreading of the real-space density even when the system is in the fully delocalized phase ($2T-\Lambda \approx 9.5$ with $\Lambda \approx 0.8$). In strong contrast, given an initial state without coherence, i.e. $|  \psi_0\rangle \sim \prod_i (b_i^{\dagger})^{n_i} | 0\rangle   $, a fast, ballistic spreading is observed on a much shorter time scale at the same choice of $T,\Lambda$ (see Fig.\ref{fig:SI_bec_mott}(b)). Note that the boson occupation number $n_i$ is chosen so that the initial density profile is the same as the one in Fig.\ref{fig:SI_bec_mott}(a). These data indicate the importance of detailed knowledge of the coherence properties of the initial state for any exact numerical simulation of our experiment or similar experiments. Furthermore, the results suggest that a coherent BEC initial state suffers a larger finite-time effect so that even a fully delocalized phase may appear to exhibit sub-ballistic transport on some short time scale. Such a finite-time effect likely does contributes to the sub-ballistic transport exponents observed in the main text Fig.~\ref{fig:largeTPhaseMap}c.

\subsubsection{Interactions}
Here we estimate the interaction energy present in the experiment in order to enable comparison of our results to previous work on subdiffusive evolution. $^{84}$Sr has a scattering length of $a = 123 a_0$, where $a_0$ is the Bohr radius. The interaction energy per particle at a typical site can then be estimated as $E_\mathrm{int} = g \bar {N} /2 \cdot \int \vert w(x) \vert ^4 \, \mathrm{d}x$. Here, $g = 4\pi\hbar^2 a / m$ is the strength of contact interaction, $\bar N$ is the mean number of particles in a single lattice site, and $w(x)$ is the ground-state Wannier function localized in a single site. We then further approximate each lattice site as a harmonic oscillator and use its Gaussian ground-state wave function for $w(x)$. From images of our atomic distribution, we estimate that in the deeply localized regime $\bar{N} \simeq 500$ on the large $T$, $\Lambda$ phase diagram, and $\bar{N} \simeq 900$ on the small $T$, $\Lambda$ phase diagram. Using the larger $\bar N = 900$, our estimates give $E_\mathrm{int}\approx 7.45 J$ for our system and thus $U = 2 E_\mathrm{int} / \bar{N} = 0.016 J$. Previous works have have shown that interaction energies on this scale give rise to subdiffusive transport exponents even in the deeply localized regime \cite{lucioni2011observation}. This provides a possible explanation for the observed nonvanishing transport exponent in the corresponding experimental regime. The same works have also investigated contributions to subdiffusive transport due to radial excitations resulting from weak radial confinement of atoms in the optical lattice, which may play a role here as well.

\subsection{Fraction of multifractal states and dynamical transport} \label{sec:multifractal_fraction}

\subsubsection{Fraction of multifractal states}
\begin{figure}[htb]
	\centering
		\includegraphics[width=0.8\textwidth]{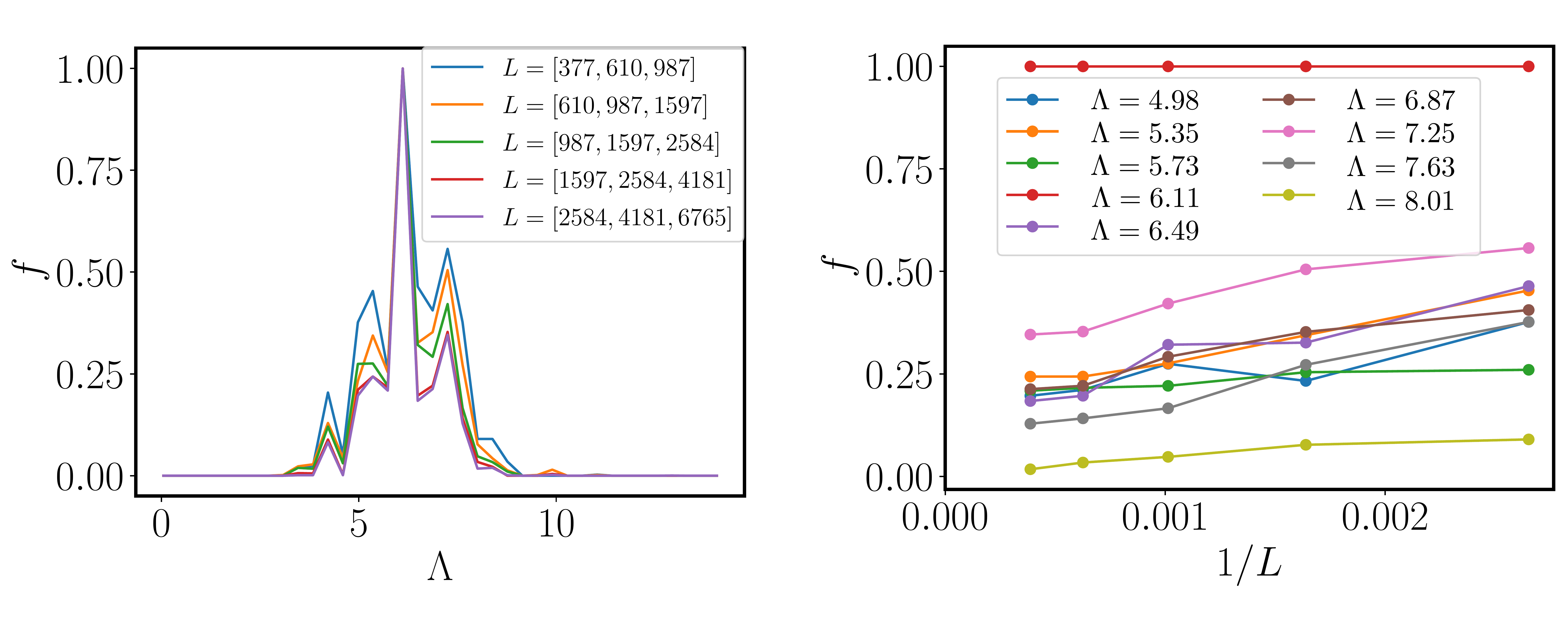}
	\caption{Left: number fraction $f$ of multifractal states, varying $\Lambda$ and $T$ along the solid line plotted in Figure~4a of the main text.  Right: Finite-size scaling of $f$ in $1/L$ at various values of $\Lambda$. System sizes are chosen from Fibonacci numbers and the wave number is chosen as $\alpha=F_{i-1}/F_i$, i.e. the ratio of two consecutive Fibonacci numbers, so that the exact duality holds for the Floquet unitary. 
		}  \label{fig:fraction_scaling}
\end{figure}

In Fig.~\ref{fig:fraction_scaling} we present numerical data on the number fraction $f$ of multifractal states, attained by numerically diagonalizing the single-particle Floquet unitary along the line  $T = - a \Lambda +b$ with $a\approx 0.393$ and $b\approx 5.461$ (i.e. following the parameter choices in Fig.4c and Fig.4d). Here the multifractal states are identified as those states with IPR exponent $\xi$ in the interval $0.1< \xi< 0.9 $ via the scaling form $\textrm{IPR} \sim L^{-\xi}$, and the different colors denote the exponents $\xi$ extracted from different selected $L$. At the self-dual point, i.e. $\Lambda \approx 6.11$, we find that $f$ remains equal to 1 as $L$ increases,  consistent with the expectation that all single-particle states are multifractal. On the other hand, $f=0$ in the regime $\Lambda\ll 6.11$ ($\Lambda\gg 6.11)$, where all states are delocalized (localized). In the intermediate regime, we find a non-zero fraction $f$ of multifractal states for a broad range of parameters, but observe that $f$ tends to decrease as $L$ increases. This trend is also visible in the finite-size scaling of $f$ with $1/L$. However, these numerical data cannot determine whether the fraction $f$ approaches zero for parameter choices away from the self-dual line in the thermodynamic limit. 

\subsubsection{Dynamical transport}
\label{sec:dynamics}

\begin{figure}[htb]
	\centering
		\includegraphics[width=0.9\textwidth]{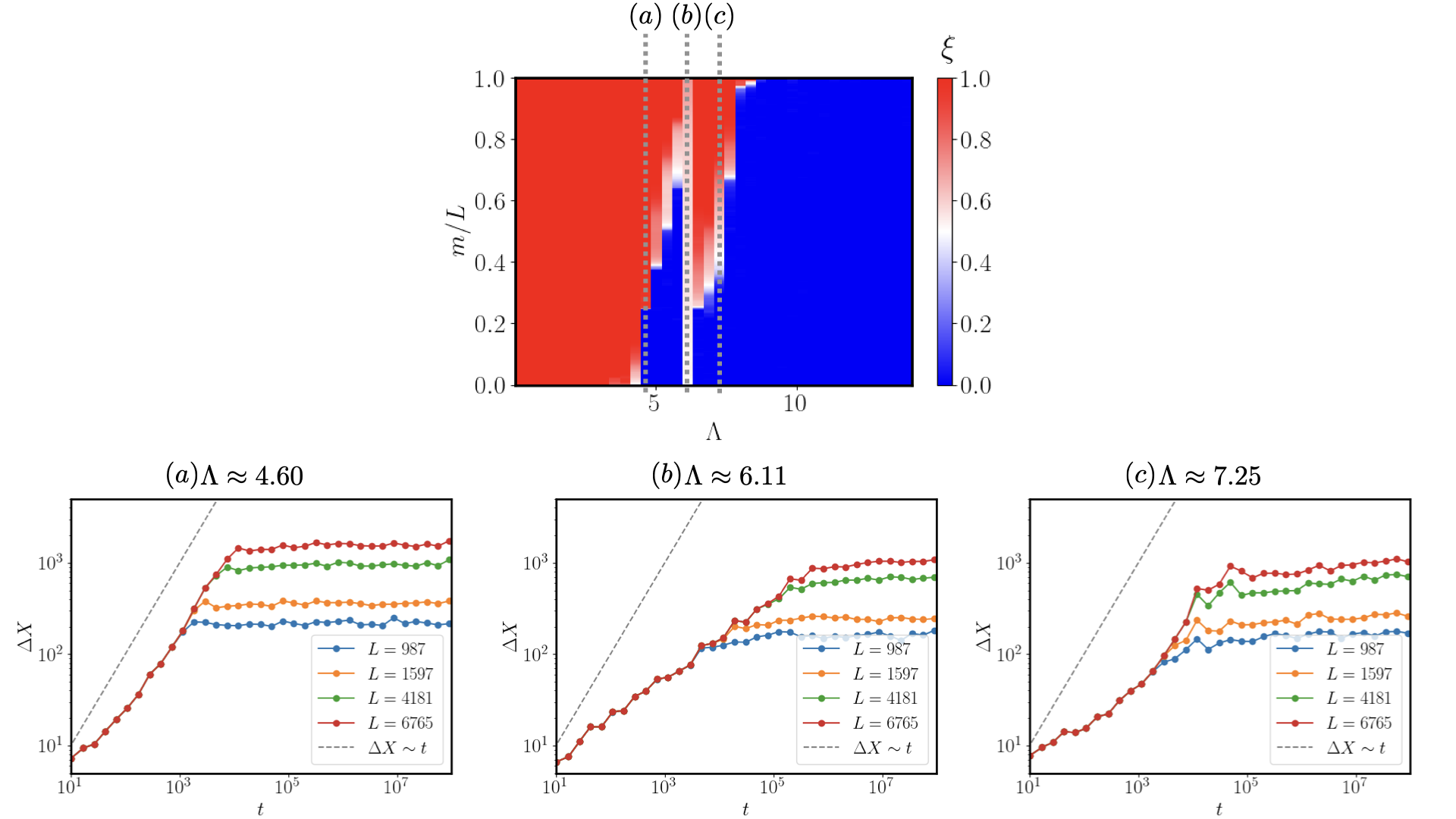}
	\caption{Top: IPR exponents $\xi$ of all single-particle states along the line $T=-a\Lambda +b $. Bottom: Spreading of a wave packet initially localized at a single lattice site for various $\Lambda, T$. Various system sizes are chosen from Fibonacci numbers and the wave number is chosen as $\alpha=F_{i-1}/F_i$, i.e. the ratio of two consecutive Fibonacci numbers.} \label{fig:fib_transport}
\end{figure}

In Fig.~\ref{fig:fib_transport} we present numerical data on the spreading of an initially localized wave packet for $T, \lambda$ along the same line $T= -a\Lambda +b$ as discussed in the previous section.   At the self-dual point, i.e. $\Lambda \approx 6.11$, all single-particle states have multi-fractal features with IPR exponents $0.1< \xi < 0.9$. Correspondingly, we observe sub-ballistic transport $\Delta X \sim t^{ \gamma}$ ($\gamma <1 $) before long-time saturation due to finite-size effects. On the other hand, at $\Lambda \approx 7.25$, where a finite fraction of multifractal states coexist with localized and delocalized states, $\Delta X$ appears to exhibit sub-ballistic transport up to a crossover time scale $t_c \sim O(10^3)$, after which the spreading becomes ballistic. To confirm the connection between this long crossover time scale and the presence of a finite fraction of multifractal states, we next focus on  the transport at $\Lambda \approx  4.6$, where the spectrum mainly consists of localized and delocalized states. In this case, we find the crossover time scale is $t_c = O(10)$. These results suggest that at the time scale and system size relevant to our experiment, the presence of a finite fraction of multifractal states is crucial to explaining the observation of sub-ballistic transport.  

\subsubsection{Comparision to static AAH model}

It is instructive to contrast this state of affairs with the (unkicked) AAH Hamiltonian which, in the thermodynamic limit, hosts multifractal states only at the delocalization-localization transition. In this case, in line with standard cross-over phenomena associated with critical phenomena, on a finite size system of O($10^3$) sites, the fraction of multifractal states is tiny when one is even a few percent away from the critical point (in non-dimensional units), and furthermore, this fraction monotonically decreases as one moves away from the critical point. In contrast, in our kAAH model, not only can one find an O(1) fraction of multifractal states on fairly large systems ($L \approx 10^3$) even when one is O(1) away from the duality line, the fraction of multifractal states does not monotonically decrease as one moves away from the duality line. This indicates that the origin of these multifractal states on finite size systems is not simply due to the proximity to the duality line, but results from the intricate nested pattern of localized and delocalized states away from the duality line \cite{ketzmerick1999efficient, prosen2001dimer}. This contrast between the unkicked and the kicked model is depicted in Fig.~\ref{fig:expt}c of the main text. Consistent with this observation is the fact that, as shown in Fig.~\ref{fig:expt}b and \ref{fig:expt}c, the extended multifractal regime in the kAAH model exists only when $T, \Lambda \gtrsim 1$. This is because when $T, \Lambda \ll 1$, the kAAH model is equivalent to a static AAH model~\cite{Qin_Yin_Chen_2014,cadez_Mondaini_Sacramento_2017}. Longer kick periods allow tunneling between kicks, endowing the effective Hamiltonian with higher-order longer-range hopping terms. We note in this connection that it has been previously argued that long-range hopping can stabilize a multifractal phase in the static AAH model~\cite{deng2019one} and that an extended multifractal phase has been also argued to exist in other Floquet quasiperiodic models~\cite{Moessner_fractal_2018, Sengupta_fractal_2021, lu2021spacetime, liu2022anomalous}, although we are not aware of detailed finite-size scaling of multifractal states in these models akin to those for the kAAH model~\cite{ketzmerick1999efficient, prosen2001dimer}. 

\section{Data Analysis}\label{SI-sec:data-analysis}

The column-integrated atomic  density  is obtained via standard calibrated absorption imaging. We perform fringe removal and background subtraction on the resulting images, then integrate the density along the axis perpendicular to the optical lattice. The resulting one-dimensional density profile is fitted to a Gaussian, and the width and center position are extracted from the fit.

\subsection{Data cutting procedure}
\label{SI-subsec:data-cuts}

\begin{figure}[ht]
\includegraphics[scale=1]{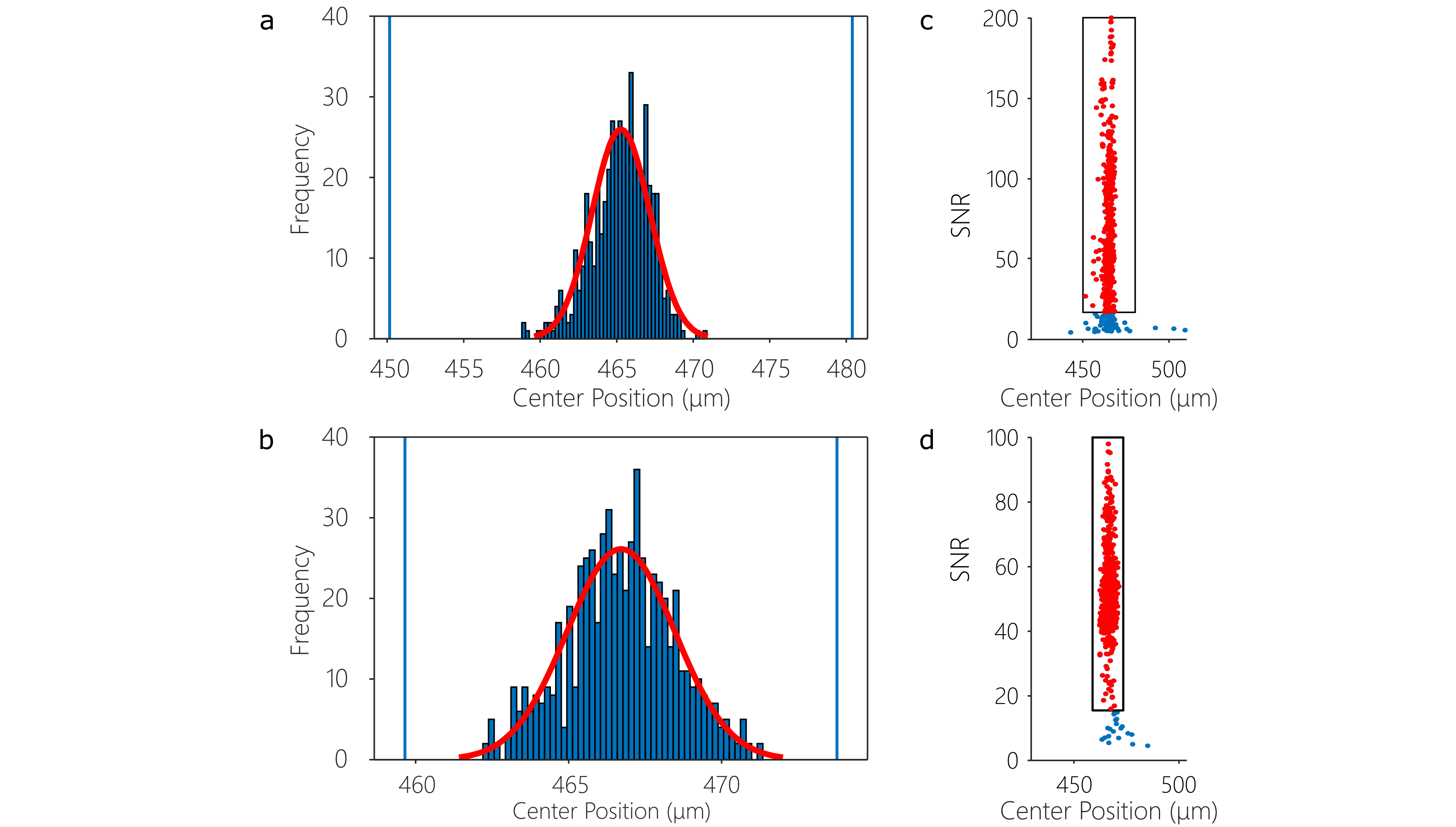}
    \caption{A depiction of data cutting procedures. \textbf{a},\textbf{c}, Cuts were made for data wherein the center of the cloud deviated by more than $N$ standard deviations from the mean. \textbf{a}, \textbf{b}, Data representing the small $\Lambda$, small $T$ phase diagram data in Fig.~\ref{fig:2DPhaseDiagram}c, $N = 2$. \textbf{c}, \textbf{d}, Data representing the large $\Lambda$, $T$ phase diagram data in Fig.~\ref{fig:largeTPhaseMap}a, $N = 4$. An additional cut was made on SNR, computed as the ratio of our signal amplitude to the standard deviation in optical density of a region without atoms. Data with $\text{SNR} < 0.4 \times \overline{\text{SNR}}$ was cut.}\label{fig:cut_explain}   
\end{figure}

In parameter regimes where the frequency content of the kicking waveform drives excitations to higher bands  (as in the crosshatched upper-right corner of Fig.\ref{fig:2DPhaseDiagram}d) or higher transverse modes (as in the crosshatched horizontal bands on Fig.\ref{fig:largeTPhaseMap}b), the atoms tend to rapidly spread out and their optical density becomes too low for reliable fitting in the presence of typical experimental noise. Additionally, large center-of-mass transport may occur in these regimes as the heated atoms ``slosh'' back and forth. As discussed in the main text, any such heating breaks the mapping of the experimental Hamiltonian to the tight-binding kAAH model. To diagnose points at which this occurs and deterministically exclude them from the associated phase diagram or transport measurement, we have made cuts based on the center position of the fit and a signal-to-noise ratio (SNR) extracted from the original density images. The distribution of center positions across all runs in a given data set are fitted to a Gaussian, and data is excluded if its center position differs from the mean by $N$ standard deviations. See Fig.~\ref{fig:cut_explain} for a visualization of this process. Additionally, to account for cases where poor fitting happened to yield a center position close to the mean and so make the first cut, we introduced a cut based on SNR. The SNR was computed for each image as the ratio of the noise level --- computed as the standard deviation of optical density within a region of the image known not to contain atoms --- to the signal amplitude near the OD peak. We omit data with $\text{SNR} < 0.4 \times \bar{\text{SNR}}$ where $\bar{\text{SNR}}$ is the mean value.

\subsection{Extraction of Transport Exponents}\label{SI-subsec:transport-exponents}

\begin{figure*}[ht]
\includegraphics[scale=1.08]{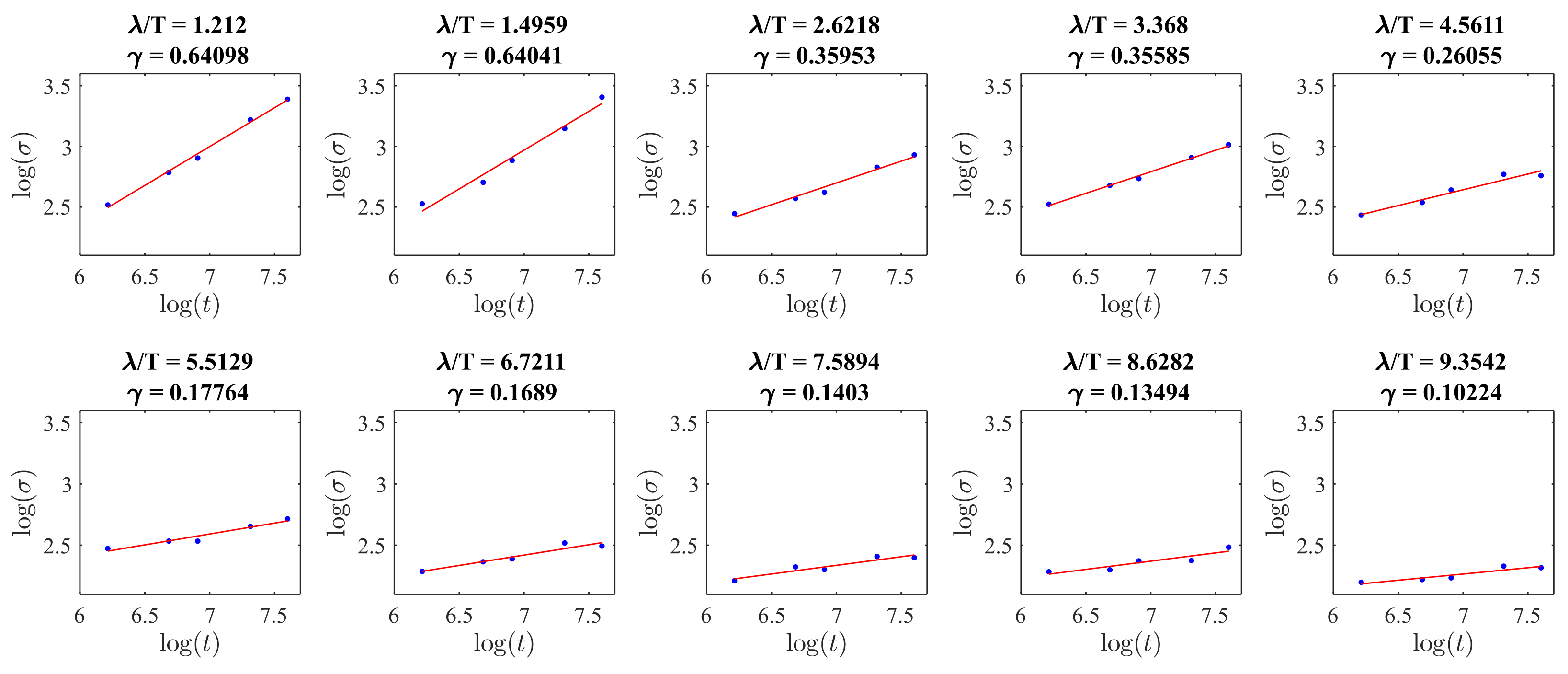}
    \caption{Width $\sigma_x(t)$ vs. time $t$ data in the small T, $\Lambda$ regime used for extracting the transport exponent $\gamma$ via fitting to a power law: $\sigma_x(t) \propto t^\gamma$. Data was taken as a horizontal cut (varying $\Lambda$) on the phase diagram shown in Fig~\ref{fig:smallPDtransport}d. The width corresponding to the midpoint of the transition of transport exponents from localized to delocalized was chosen as our colormap center (white) for the phase maps.}
    \label{fig:smallPD-transport-fits}
\end{figure*}

\begin{figure*}[ht]
\includegraphics[scale=1.48]{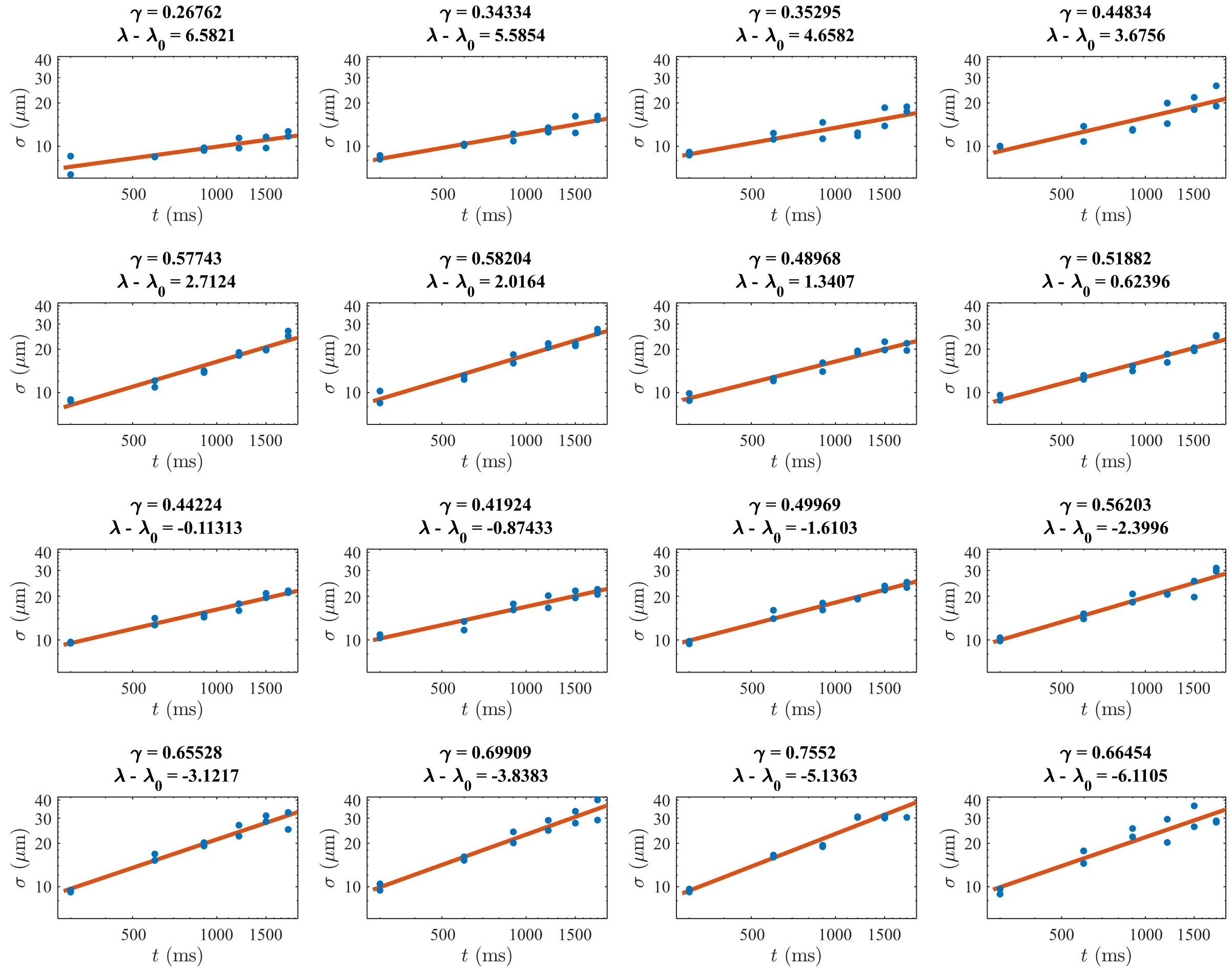}
\caption{Width $\sigma_x(t)$ vs. time $t$ data used for extracting the transport exponent $\gamma$ via fitting to a power law: $\sigma_x(t) \propto t^\gamma$. The $\sigma_x(t)$ datasets were taken as points along a nearly transverse line to the $\Lambda = 2T$ static-AA transition line. Extracted transport exponents are plotted in Fig.~\ref{fig:largeTPhaseMap}c.}
\label{fig:largePD-transport-fits}
\end{figure*}

In order to extract transport exponents, the time evolution of the atom distribution width $\sigma_x$ was treated via two methods. The first is the solution to a generalized diffusion equation, as has been studied in the context of static quasicrystals \cite{lucioni2011observation}. The generalized diffusion equation is
    \begin{align}
        \frac{ \partial \sigma_x }{\partial t }
        &= C \sigma_x^{-p}
        \quad
        \underset{\sigma_x(t = 0) = \sigma_0}{\implies}
        \quad
        \sigma_x(t) = \sigma_0 \left( 1 + \frac{t}{t_0} \right)^\gamma, \label{lucioni-form}
    \end{align}
where $\gamma = 1/(1+p)$ is the transport exponent, $C$ is a modified diffusion constant, and $t_0 = C / (\gamma \sigma_0^{1/\gamma})$ is the ``turn-on" time for the asymptotic regime. This form takes account of the density profile's initial width $\sigma_0$, allowing $\gamma$ to be extracted from short and intermediate time transport behavior. Note that for evolution times $t$ much greater than the ``turn-on" time $t_0$, one recovers the expected power-law scaling: $\lim\limits_{t/t_0 \gg 1} \sigma_x(t) \propto t^\gamma$.
Fitting to equation \ref{lucioni-form} was used in the low $T$ and $\Lambda$ regime for initial observation of a localization transition as a sudden change in the transport exponent $\gamma$. These observations were used to calibrate the phase diagram colormap, the neutral center color of which is set to this transition point. Examples of fitting experimental data to equation \ref{lucioni-form} are shown in Fig.~\ref{fig:2DPhaseDiagram}b. For the transport measurements reported in Fig.~\ref{fig:largeTPhaseMap}c, the width evolution was instead fit to a power law at intermediate and late times, $\sigma_x(t) \propto t^\gamma$. In this regime, difficulty resolving the expansion at short times led to a difficulty in the simultaneous constraint of $t_0$ and $\gamma$ when fitting to equation~\eqref{lucioni-form}, and so simple power-law fitting with one less free parameter proved to be more reliable. It should be noted that due to experimental limitations on the maximum duration of a single run, our width expansion data likely lies in an intermediate regime between short and long time behavior -- this is confirmed by the fact that our experiment times are not large compared to extracted $t_0$ values, and that we observe uncharacteristically low transport exponents even deep into the delocalized regime. For this reason, we should expect that a simple power-law fit to even our latest time data will yield transport exponents which are lower than would be observed at longer times. This may account for the observation of transport exponents $\gamma < 1$ even in the deeply localized regime, as seen in Fig.~\ref{fig:largeTPhaseMap}c. The width vs. time data and fits used to extract the transport exponents $\gamma$ are provided in figures Fig.~\ref{fig:smallPD-transport-fits} (for the small $\Lambda$, $T$ regime) and Fig.~\ref{fig:largePD-transport-fits} (for the large $\Lambda$, $T$ regime).  Error bars for fit parameters are extracted as 95\% confidence intervals from the fitting algorithm.

\end{document}